\documentclass[sigconf]{acmart}
\def\BibTeX{{\rm B\kern-.05em{\sc i\kern-.025em b}\kern-.08emT\kern-.1667em\lower.7ex\hbox{E}\kern-.125emX}}   
\setcopyright{acmcopyright}
\copyrightyear{2019} 
\acmYear{2019} 
\acmConference[MM '19]{Proceedings of the 27th ACM International Conference on Multimedia}{October 21--25, 2019}{Nice, France}
\acmBooktitle{Proceedings of the 27th ACM International Conference on Multimedia (MM '19), October 21--25, 2019, Nice, France}
\acmPrice{15.00}
\acmDOI{10.1145/3343031.3351028}
\acmISBN{978-1-4503-6889-6/19/10}

\acmSubmissionID{44}

\usepackage{booktabs}   %
\usepackage{multirow,tabularx}  %
\usepackage{bm}  %
\usepackage[caption=false,font=footnotesize]{subfig}  %
\begin{document}
\fancyhead{}

%
\title[Quality Assessment of In-the-Wild Videos]{Quality Assessment of In-the-Wild Videos}

%
 
 \author{Dingquan Li}
 \email{dingquanli@pku.edu.cn}
 \orcid{0000-0002-5549-9027}
 \affiliation{%
  \institution{NELVT, LMAM, School of Mathematical Sciences \& BICMR, Peking University}
 }
 
 \author{Tingting Jiang}
 \email{ttjiang@pku.edu.cn}
 \affiliation{%
  \institution{NELVT, Department of Computer Science, Peking University}
 }

 \author{Ming Jiang}
 \email{ming-jiang@pku.edu.cn}
 \affiliation{%
  \institution{NELVT, LMAM, School of Mathematical Sciences \& BICMR, Peking University}
 }

\renewcommand{\shortauthors}{D. Li, T. Jiang, and M. Jiang}

%
\begin{abstract}
Quality assessment of in-the-wild videos is a challenging problem because of the absence of reference videos and shooting distortions. 
Knowledge of the human visual system can help establish methods for objective quality assessment of in-the-wild videos.
In this work, we show two eminent effects of the human visual system, namely,  content-dependency and temporal-memory effects, could be used for this purpose. 
We propose an objective no-reference video quality assessment method by integrating both effects into a deep neural network.  
For content-dependency, we extract features from a pre-trained image classification neural network for its inherent content-aware property. 
For temporal-memory effects, long-term dependencies, especially the temporal hysteresis, are integrated into the network with a gated recurrent unit and a subjectively-inspired temporal pooling layer.
To validate the performance of our method, experiments are conducted on three publicly available in-the-wild video quality assessment databases: KoNViD-1k, CVD2014, and LIVE-Qualcomm, respectively. 
Experimental results demonstrate that our proposed method outperforms five state-of-the-art methods by a large margin, specifically, 12.39\%, 15.71\%, 15.45\%, and 18.09\% overall performance improvements over the second-best method VBLIINDS, in terms of SROCC, KROCC, PLCC  and RMSE, respectively. 
Moreover, the ablation study verifies the crucial role of both the content-aware features and the modeling of temporal-memory effects. The PyTorch implementation of our method is released at \url{https://github.com/lidq92/VSFA}.
\end{abstract}

%



%

%
%
\maketitle

\section{Introduction}
\label{sec:introduction}
Nowadays, most videos are captured in the wild by users with diverse portable mobile devices, which may contain annoying distortions due to out of focus, object motion, camera shake, or under/over exposure. 
Thus, it is highly desirable to automatically identify and cull low-quality videos, prevent their occurrence by quality monitoring processes during acquisition, or repair/enhance them with the quality-aware loss.
To achieve this goal, quality assessment of in-the-wild videos is a precondition.
However, this is a challenging problem due to the fact that the ``perfect'' source videos are not available and the shooting distortions are unknown. 
There is an essential difference between in-the-wild videos and synthetically-distorted videos, \textit{i.e.}, the former contains a mass of content and may suffer from complex mixed real-world distortions that are temporally heterogeneous.
On account of this, current state-of-the-art video quality assessment (VQA) methods (\textit{e.g.}, VBLIINDS~\cite{saad2014blind} and VIIDEO~\cite{mittal2016completely}) validated on traditional synthetic VQA databases~\cite{seshadrinathan2010study, moorthy2012video} fail in predicting the quality of in-the-wild videos~\cite{men2017empirical,ghadiyaram2018capture, nuutinen2016cvd2014,sinno2018large}. 

This work focuses on the problem ``quality assessment of in-the-wild videos''. 
Since humans are the end-users, we believe that knowledge of the human visual system (HVS) can help establish objective methods for our problem. 
Specifically, two eminent effects of HVS are incorporated into our method.

\begin{figure}
\begin{center}
  \includegraphics[width=.9\columnwidth]{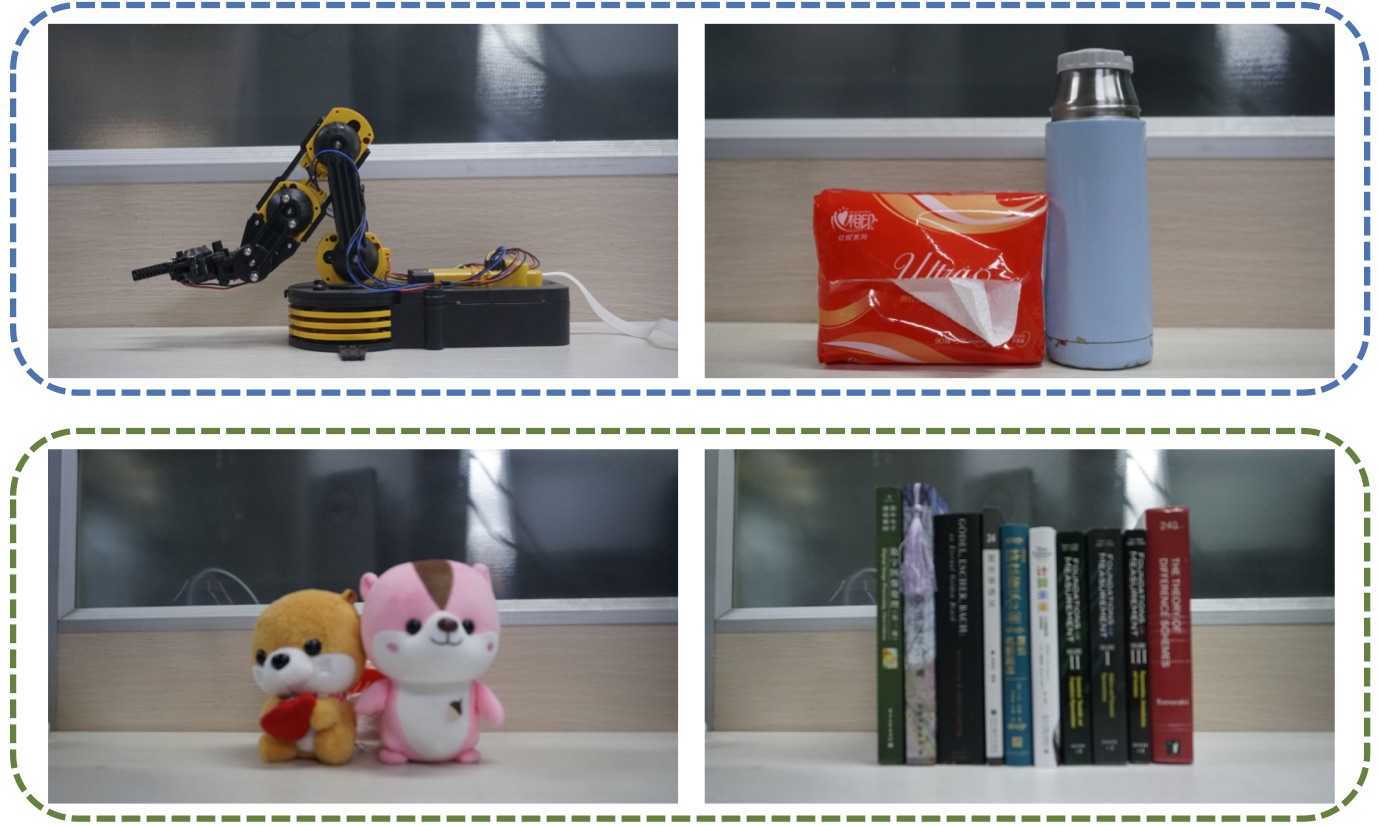}%
\label{fig:P1}
\end{center}
  \caption{[Best viewed when zoomed in] Human judgments of visual quality are content-dependent. The first/second row shows a pair of in-focus/out-of-focus images. Every two images in a pair are taken in the same shooting condition, and they only differ in image content. However, user study shows that humans consistently prefer the left ones.} 
  \Description{}
  \label{fig:video content}
\end{figure}

\textbf{Human judgments of visual image/video quality depend on content}, which is well known in many subjective experiments~\cite{siahaan2018semantic,triantaphillidou2007image,wang2017videoset,bampis2017study,duanmu2017quality,zhang2018unreasonable,mirkovic2014evaluating}. 
For images, Siahaan \textit{et al.} show that scene and object categories influence human judgments of visual quality for JPEG compressed and blurred images~\cite{siahaan2018semantic}.
Two compressed images with the same compression ratio may have different subjective quality if they contain different scenes~\cite{triantaphillidou2007image}, since the scene content can have different impact on the compression operations and the visibility of artifacts. 
For videos, similar content dependency can be found in compressed video quality assessment~\cite{wang2017videoset,mirkovic2014evaluating} and quality-of-experience of streaming videos~\cite{bampis2017study,duanmu2017quality}.
Unlike quality assessment of synthetically-distorted images/videos, quality assessment of in-the-wild images/videos essentially requires to compare cross-content image/video pairs (\textit{i.e.}, the pair from different reference images/videos)~\cite{mikhailiuk2018psychometric}, which may be more strongly affected by content.
To verify the correctness of this effect on our problem, we collect data and conduct a user study. 
We ask 10 human subjects to do the cross-content pairwise comparison for 201 image pairs. 
More than 7 of 10 subjects prefer one image to the other image in 82 image pairs.
For illustration, two pairs of in-the-wild images are shown in~Figure~\ref{fig:video content}.
Each image pair is taken in the same shooting conditions (\textit{e.g.}, focus length, object distance).
For the in-focus image pair in the first row, 9 of 10 subjects prefer the left one. 
For the out-of-focus image pair in the second row, 8 of 10 subjects prefer the left one to the right one.  
The only difference within a pair is the image content, so from our user study, we can infer that image content can affect human perception on quality assessment of in-the-wild images. 
We also conduct a user study for 43 video pairs, where every two videos in a pair are taken in similar settings. Similar results are found that video content could have impacts on judgments of visual quality for in-the-wild videos. 
In the supplemental material, we provide a video pair, for which all 10 subjects prefer the same video.
Thus, we consider content-aware features in our problem to address the content dependency.

\textbf{Human judgments of video quality are affected by their temporal memory}. 
Temporal-memory effects indicate that human judgments of current frame rely on the current frame and information from previous frames. And this implies that long-term dependencies exist in the VQA problem. 
More specifically, humans remember poor quality frames in the past and lower the perceived quality scores for following frames, even when the frame quality has returned to acceptable levels~\cite{seshadrinathan2011temporal}. 
This is called the temporal hysteresis effect.
It indicates that the simple average pooling strategy overestimates the quality of videos with fluctuating frame-wise quality scores. 
Since the in-the-wild video contains more temporally-heterogeneous distortions than the synthetically-distorted video, human judgments of its visual quality reflect stronger hysteresis effects.
Therefore, in our problem, modeling of temporal-memory effects should be taken into account.
 
In light of the two effects, we propose a simple yet effective no-reference (NR) VQA method with content-aware features and modeling of temporal-memory effects.
To begin with, our method extracts content-aware features from deep convolutional neural networks (CNN) pre-trained on image classification tasks, for they are able to discriminate abundant content information. 
After that, it includes a gated recurrent unit (GRU) for  modeling long-term dependencies and predicting frame quality.
Finally, to take the temporal hysteresis effects into account, we introduce a differentiable subjectively-inspired temporal pooling model, and embed it as a layer into the network to output the overall video quality. 

To demonstrate the performance of our method, we conduct experiments on three publicly available databases, \textit{i.e.}, KoNViD-1k~\cite{hosu2017konstanz}, LIVE-Qualcomm~\cite{ghadiyaram2018capture} and CVD2014~\cite{nuutinen2016cvd2014}. 
Our method is compared with five state-of-the-art methods, and its superior performance is proved by the experimental results. 
Moreover, the ablation study verifies the key role of each component in our method. 
This suggests that incorporating the knowledge of HVS could make objective methods more consistent with human perception.

The main contributions of this work are as follows:

\begin{itemize}
\item An objective NR-VQA method and the first deep learning-based model is proposed for in-the-wild videos.
\item To our best knowledge, it is the first time that a GRU network is applied to model the long-term dependencies for quality assessment of in-the-wild videos and a differentiable temporal pooling model is put forward to account for the hysteresis effect.
\item The proposed method outperforms the state-of-the-art methods by large margins, which is demonstrated by experiments on three large-scale in-the-wild VQA databases.
\end{itemize}

\begin{figure*}[!htb]
    \centering
    \includegraphics[width=.8\textwidth]{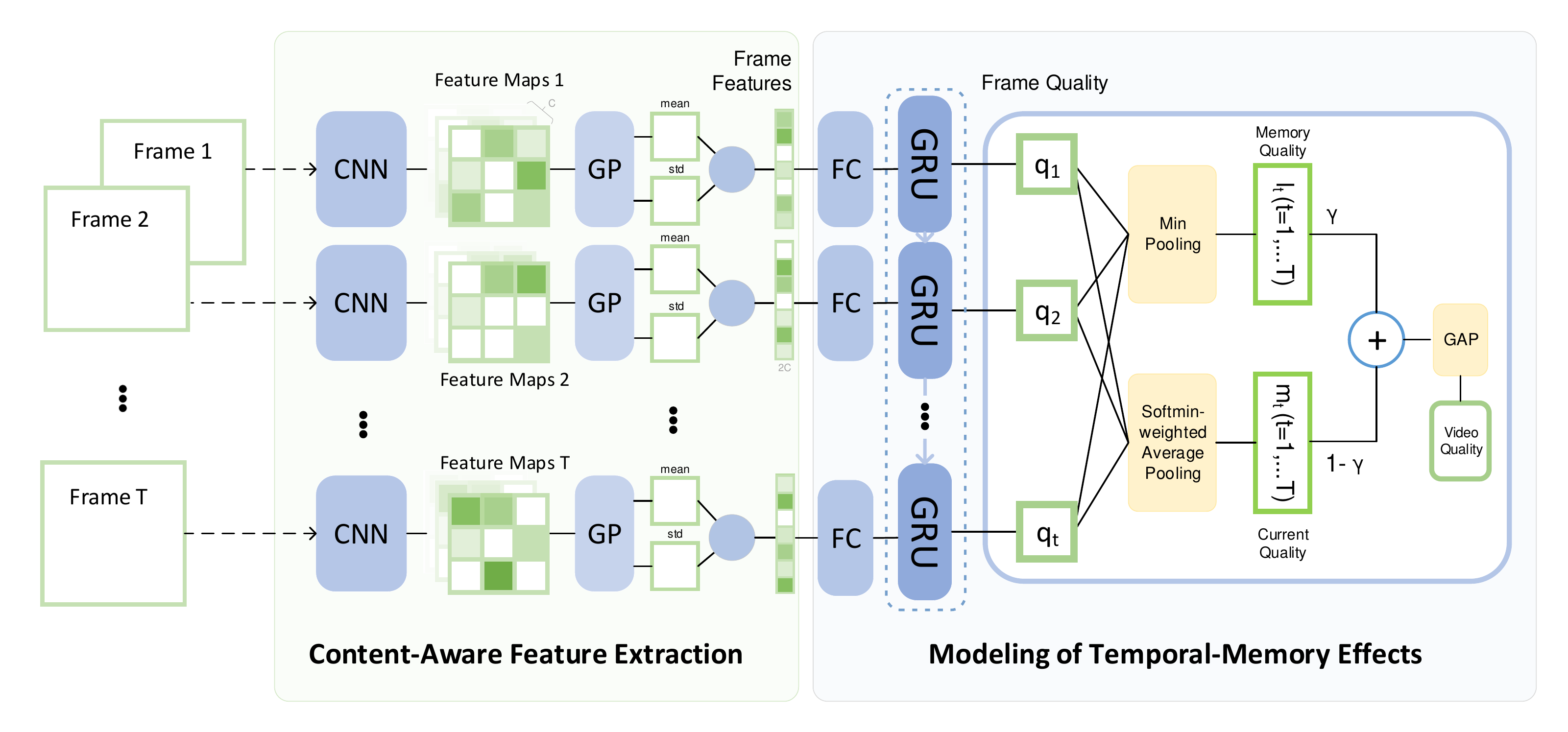}
    \caption{The overall framework of the proposed method. It mainly consists of two modules. The first module ``content-aware feature extraction'' is a pre-trained CNN with effective global pooling (GP) serving as a feature extractor. The second module ``modeling of temporal-memory effects'' includes two sub-modules: one is a GRU network for modeling long-term dependencies; the other is a subjectively-inspired temporal pooling layer accounting for the temporal hysteresis effects. Note that the GRU network is the unrolled version of one GRU and the parallel CNNs/FCs share weights.}
    \label{fig:framework}
\end{figure*}

\section{Related Work}
\label{sec:related}
\subsection{Video Quality Assessment}
Traditional VQA methods consider structures ~\cite{wang2004video,wang2012novel}, gradients~\cite{lu2019spatiotemporal}, motion~\cite{seshadrinathan2010motion,manasa2016optical-NR}, energy~\cite{li2016spatiotemporal}, saliency~\cite{zhang2017study, you2014attention}, or natural video statistics~\cite{ghadiyaram2017no,zhu2017blind,mittal2016completely,saad2014blind}.
Besides, quality assessment can be achieved by fusion of primary features~\cite{freitas2018using,li2016no}. 
Recently, four deep learning-based VQA methods are proposed~\cite{zhang2018blind,liu2018end,kim2018deep,zhang2019objective}.
Kim~\textit{et al.}~\cite{kim2018deep} utilize CNN models to learn the spatio-temporal sensitivity maps. 
Liu~\textit{et al.}~\cite{liu2018end} exploit the 3D-CNN model for codec classification and quality assessment of compressed videos. 
Zhang~\textit{et al.}~\cite{zhang2018blind,zhang2019objective} apply the transfer learning technique with CNN for video quality assessment.
However, all these methods are trained, validated, and tested on synthetically distorted videos. Streaming video quality-of-experience is relevant to video quality but beyond the scope of this paper, and an interested reader can refer to the good surveys~\cite{seufert2014survey,juluri2015measurement}.

Quality assessment of in-the-wild videos is a quite new topic in recent years~\cite{hosu2017konstanz,ghadiyaram2018capture, nuutinen2016cvd2014,sinno2018large}. 
Four relevant databases have been constructed and corresponding subjective studies have been conducted. 
Overall, CVD2014~\cite{nuutinen2016cvd2014}, KoNViD-1k~\cite{hosu2017konstanz}, and LIVE-Qualcomm~\cite{ghadiyaram2018capture} are publicly available, while LIVE-VQC~\cite{sinno2018large} will be available soon.
Due to the fact that we cannot access the pristine reference videos in this situation, only NR-VQA methods are applicable. 
Unfortunately, the evaluation of current state-of-the-art NR-VQA methods~\cite{mittal2016completely,saad2014blind} on these video databases shows a poor performance~\cite{men2017empirical,ghadiyaram2018capture, nuutinen2016cvd2014,sinno2018large}. 
Existing deep learning-based VQA models are unfeasible in our problem since they either need the reference information~\cite{zhang2018blind,kim2018deep,zhang2019objective} or only suit for compression artifacts~\cite{liu2018end}.
Thus, this motivates us to propose the first deep learning-based model that is capable of predicting the quality of in-the-wild videos.

\subsection{Content-Aware Features}
Content-aware features can help addressing content-dependency on the predicted image/video quality, so as to improve the performance of objective models~\cite{jaramillo2016content,siahaan2018semantic,wu2019blind,li2019has}.
Jaramillo~\textit{et al.}~\cite{jaramillo2016content} extract handcrafted content-relevant features to tune existing quality measures.
Siahaan~\textit{et al.}~\cite{siahaan2018semantic} and Wu~\textit{et al.}~\cite{wu2019blind} utilize semantic information from the top layer of pre-trained image classification networks to incorporate with traditional quality features. 
Li~\textit{et al.}~\cite{li2019has} exploit the deep semantic feature aggregation of multiple patches for image quality assessment. 
It is shown that these deep semantic features alleviate the impact of content on the quality assessment task. 
Inspired by their work, we consider using pre-trained image classification networks for content-aware feature extraction as well. 
Unlike the work in~\cite{li2019has}, to get the features, we directly feed the whole frame into the network and apply not only global average pooling but also global standard deviation pooling to the output semantic feature maps. 
Since our work aims at the VQA task, we further put forward a new module for modeling temporal characteristics of human behaviors when rating video quality.

\subsection{Temporal Modeling}

The temporal modeling in the VQA field can be viewed in two aspects, \textit{i.e.}, feature aggregation and quality pooling. 

In the feature aggregation aspect, most methods aggregate frame-level features to video-level features by averaging them over the temporal axis~\cite{men2017empirical, saad2014blind, manasa2016optical-NR,men2018spatiotemporal, freitas2018using,li2016spatiotemporal}. 
Li~\textit{et al.}~\cite{li2016no} adopt a 1D convolutional neural network to aggregate the primary features for a time interval. 
Unlike the previous methods, we consider using GRU network to model the long-term dependencies for feature integration.

In the quality pooling aspect, the simple average pooling strategy is adopted by many methods~\cite{vu2014vis3,seshadrinathan2010motion, liu2018end, zhu2017blind,mittal2016completely}. 
Several pooling strategies considering the recency effect or the worst quality section influence are discussed in~\cite{rimac2009influence,seufert2013pool}.
Kim~\textit{et al.}~\cite{kim2018deep} adopt a convolutional neural aggregation network (CNAN) for learning frame weights, then the overall video quality is calculated by the weighted average of frame quality scores. 
Seshadrinathan and Bovik~\cite{seshadrinathan2011temporal} notice the temporal hysteresis effect in the subjective experiments, and propose a temporal hysteresis pooling strategy for quality assessment. 
The effectiveness of this strategy has been verified in~\cite{seshadrinathan2011temporal, xu2014no, choi2018video}. 
We also take account of the temporal hysteresis effects. 
However, the temporal pooling model in~\cite{seshadrinathan2011temporal} is not differentiable. 
So we introduce a new one with subjectively-inspired weights which can be embedded into the neural network and be trained with back propagation as well. 
In the experimental part, we will show that this new temporal pooling model with subjectively-inspired weights is better than the CNAN temporal pooling~\cite{kim2018deep} with learned weights.

\section{The Proposed Method}
\label{sec:framework}
In this section, we introduce a novel NR-VQA method by integrating knowledge of the human visual system into a deep neural network.
The framework of the proposed method is shown in Figure~\ref{fig:framework}.
It extracts content-aware features from a modified pre-trained CNN with global pooling (GP) for each video frame. 
Then the extracted frame-level features are sent to a fully-connected (FC) layer for dimensional reduction followed by a GRU network for long-term dependencies modeling. 
In the meantime, the GRU outputs the frame-wise quality scores. Lastly, to account for the temporal hysteresis effect, the overall video quality is pooled from these frame quality scores by a subjectively-inspired temporal pooling layer. We will detail each part in the following.

\subsection{Content-Aware Feature Extraction}
For in-the-wild videos, the perceived video quality strongly depends on the video content as described in Section~\ref{sec:introduction}. 
This can be attributed to the fact that, the complexity of distortions, the human tolerance thresholds for distortions, and the human preferences could vary for different video content/scenes.

To evaluate the perceived quality of in-the-wild videos, the above observation motivates us to extract features that are not only perceptual (distortion-sensitive) but also content-aware.
The image classification models pre-trained on ImageNet~\cite{deng2009imagenet} using CNN possess the discriminatory power of different content information.
Thus, the deep features extracted from these models (\textit{e.g.}~ResNet~\cite{he2016deep}) are expected to be content-aware. 
Meanwhile, the deep features are distortion-sensitive~\cite{dodge2016understanding}. 
So it is reasonable to extract content-aware perceptual features from pre-trained image classification models.

Firstly, assuming the video has $T$ frames, we feed the video frame $\mathbf{I}_t (t=1,2,\dots,T)$ into a pre-trained CNN model and output the deep semantic feature maps $\mathbf{M}_t$ from its top convolutional layer:
\begin{equation}\label{eq:CNN}
\mathbf{M}_t = \mathrm{CNN}(\mathbf{I}_t).
\end{equation}

$\mathbf{M}_{t}$ contains a total of $C$ feature maps.
Then, we apply spatial GP for each feature map of $\mathbf{M}_{t}$. 
Applying the spatial global average pooling operation ($\mathrm{GP}_{\mathrm{mean}}$) to $\mathbf{M}_t$ discards much information of $\mathbf{M}_t$. 
We further consider the spatial global standard deviation pooling operation ($\mathrm{GP}_{\mathrm{std}}$) to preserve the variation information in $\mathbf{M}_t$. 
The output feature vectors of $\mathrm{GP}_{\mathrm{mean}}, \mathrm{GP}_{\mathrm{std}}$ are $\mathbf{f}_{t}^{\mathrm{mean}}, \mathbf{f}_{t}^{\mathrm{std}}$ respectively.
\begin{equation}\label{eq:GP}
\begin{aligned}
\mathbf{f}_{t}^{\mathrm{mean}} & = \mathrm{GP}_{\mathrm{mean}}(\mathbf{M}_t),\\
\mathbf{f}_{t}^{\mathrm{std}} & = \mathrm{GP}_{\mathrm{std}}(\mathbf{M}_t).\\
\end{aligned}
\end{equation}

After that, $\mathbf{f}_{t}^{\mathrm{mean}}$ and $\mathbf{f}_{t}^{\mathrm{std}}$ are concatenated to serve as the content-aware perceptual features $\mathbf{f}_t$:
\begin{equation}\label{eq:concatenation}
\mathbf{f}_t = \mathbf{f}_{t}^{\mathrm{mean}}\oplus\mathbf{f}_{t}^{\mathrm{std}},
\end{equation}
where \(\oplus\) is the concatenation operator and the length of $\mathbf{f}_t$ is $2C$.

\subsection{Modeling of Temporal-Memory Effects}

Temporal modeling is another important clue for designing objective VQA models. We model the temporal-memory effects in two aspects. In the feature integration aspect, we adopt a GRU network for modeling the long-term dependencies in our method. In the quality pooling aspect, we propose a subjectively-inspired temporal pooling model and embed it into the network.

\textbf{Long-term dependencies modeling}.
Existing NR-VQA methods cannot well model the long-term dependencies in the VQA task. 
To handle this issue, we resort to GRU~\cite{cho2014learning}. It is a recurrent neural network model with gates control which is capable of both integrating features and learning long-term dependencies. 
Specifically, in this paper, we consider using GRU to integrate the content-aware perceptual features and predict the frame-wise quality scores.

The extracted content-aware features are of high dimension, which is not easy for training GRU. 
Therefore, it is better to perform dimension reduction before feeding them into GRU. 
It could be beneficial by performing dimension reduction with other steps in the optimization process jointly. 
In this regard, we perform dimension reduction using a single FC layer, that is:
\begin{equation}\label{eq:linear DR}
\mathbf{x}_t = \mathbf{W}_{fx}\mathbf{f}_t+\mathbf{b}_{fx},
\end{equation}
where $\mathbf{W}_{fx}$ and $\mathbf{b}_{fx}$ are the parameters in the single FC layer. 
Without the bias term, it acts as a linear dimension reduction model.

After dimension reduction, the reduced features $\mathbf{x}_t (t=1,\cdots, T)$ are sent to GRU. 
We consider the hidden states of GRU as the integrated features, whose initial values are $\mathbf{h}_0$. 
The current hidden state $\mathbf{h}_t$ is calculated from the current input $\mathbf{x}_t$ and the previous hidden state $\mathbf{h}_{t-1}$, that is:
\begin{equation}\label{eq:GRU}
\mathbf{h}_t = \mathrm{GRU}(\mathbf{x}_t, \mathbf{h}_{t-1}).
\end{equation}


With the integrated features $\mathbf{h}_t$, we can predict the frame quality score $q_t$ by adding a single FC layer:
\begin{equation}
q_t = \mathbf{W}_{hq}\mathbf{h}_t+\mathbf{b}_{hq},
\end{equation}
where $\mathbf{W}_{hq}$ and $\mathbf{b}_{hq}$ are the weight and bias parameters.

\textbf{Subjectively-inspired temporal pooling}.
In subjective experiments, subjects are intolerant of poor quality video events~\cite{park2013video}. 
More specifically, temporal hysteresis effect is found in the subjective experiments, \textit{i.e.}, subjects react sharply to drops in video quality and provide poor quality for such time interval, but react dully to improvements in video quality thereon~\cite{seshadrinathan2011temporal}. 

A temporal pooling model is adopted in~\cite{seshadrinathan2011temporal} to account for the hysteresis effect. 
Specifically, a memory quality element is defined as the minimum of the quality scores over the previous frames; a current quality element is defined as a sort-order-based weighted average of the quality scores over the next frames; the approximate score is calculated as the weighted average of the memory and current elements; the video quality is computed as the temporal average pooling of the approximate scores. 
However, there are some limitations on directly applying this model to the NR quality assessment of in-the-wild videos. 
First, this model requires the reliable frame quality scores as input, which cannot be provided in our task. 
Second, the model in~\cite{seshadrinathan2011temporal} is not differentiable due to the sort-order-based weights in the definition of the current quality element. Thus it cannot be embedded into the neural network. 
In our problem, since we only have access to the overall subjective video quality, we need to learn the neural network without frame-level supervision. 
Thus, to connect the predicted frame quality score $q_t$ to the video quality $Q$, we put forward a new differentiable temporal pooling model by replacing the sort-order-based weight function in~\cite{seshadrinathan2011temporal} with a differentiable weight function, and embed it into the network. 
Details are as follow.

To mimic the human's intolerance to poor quality events, we define a memory quality element $l_t$ at the $t$-th frame as the minimum of quality scores over the previous several frames:
\begin{equation}\label{eq:memory score}
\begin{aligned}
l_t & =q_t, & \mbox{for}\, \ t=1, \\
l_t & = \min_{k\in V_{prev}}{q_{k}}, & \mbox{for}\, \ t>1, 
\end{aligned}
\end{equation}
where $V_{prev}=\{\max{(1,t-\tau)},\cdots, t-2, t-1\}$ is the index set of the considered frames, and $\tau$ is a hyper-parameter relating to the temporal duration.

Accounting for the fact that subjects react sharply to the drops in quality but react dully to the improvements in quality, we construct a current quality element $m_t$ at the $t$-th frame, using the weighted quality scores over the next several frames, where larger weights are assigned for worse quality frames. Specifically, we define the weights $w_t^k$ by a differentiable softmin function (a composition of the negative linear function and the softmax function).
\begin{equation}\label{eq:current score}
\begin{aligned}
m_t & = \sum_{k\in V_{next}}q_{k}w_t^{k},\\
w_t^k & = \frac{e^{-q_k}}{\sum_{j\in V_{next}}e^{-q_j}}, k\in V_{next}, 
\end{aligned}
\end{equation}
where $V_{next}=\{t,t+1,\cdots, \min{(t+\tau,T)}\}$ is the index set of the related frames.

In the end, we approximate the subjective frame quality scores by linearly combining the memory quality and current quality elements. The overall video quality $Q$ is then calculated by temporal global average pooling (GAP) of the approximate scores:
\begin{equation}\label{eq:approximate quality}
q'_t  = \gamma l_t + (1-\gamma)m_t,
\end{equation}
\begin{equation}\label{eq:video quality}
Q  = \frac{1}{T}\sum_{t=1}^{T}q'_t,
\end{equation}
where $\gamma$ is a hyper-parameter to balance the contributions of memory and current elements to the approximate score.

Note that we model the temporal-memory effects with both a global module (\textit{i.e.}, GRU) and a local module  (\textit{i.e.}, subjectively-inspired temporal pooling with a window size of $2\tau+1$). The long-term dependency is always considered by GRU, no matter which value of $\tau$ in the temporal pooling is chosen.

\subsection{Implementation Details}
We choose ResNet-50~\cite{he2016deep} pre-trained on ImageNet~\cite{deng2009imagenet} for the content-aware feature extraction, and the feature maps are extracted from its `res5c' layer. In this instance, the dimension of $\mathbf{f}_t$ is $4096$.
The long-term dependencies part is a single FC layer that reduces the feature dimension from 4096 to 128, followed by a single-layer GRU network whose hidden size is set as 32. 
The subjectively-inspired temporal pooling layer contains two hyper-parameters, $\tau$ and $\gamma$, which are set as 12 and 0.5, respectively. 
We fix the parameters in the pre-trained ResNet-50 to ensure that the content-aware property is not altered, and we train the whole network in an end-to-end manner. 
The proposed model is implemented with PyTorch~\cite{paszke2017automatic}. The $L_1$ loss and Adam~\cite{kingma2014adam} optimizer with an initial learning rate 0.00001 and training batch size 16 are used for training our model.

\section{Experiments}
\label{sec:experiments}
We first describe the experimental settings, including the databases, compared methods and basic evaluation criteria. Next, we carry out the performance comparison and result analysis of our method with five state-of-the-art methods. After that, an ablation study is conducted. Then, we show results of different choices of feature extractor and temporal pooling strategy. Finally, the adding value of motion information and computational efficiency are discussed.
\begin{table*}[!hbt]
    \centering
    \caption{Performance comparison on the three VQA databases. Mean and standard deviation (std) of the performance values in 10 runs are reported, \textit{i.e.}, mean ($\pm$ std).  `Overall Performance' shows the weighted-average performance values over all three databases, where weights are proportional to database-sizes. In each column, the best and second-best values are marked in boldface and underlined, respectively.}
    \label{tab:performance}

    \begin{small}
    
    \resizebox{\textwidth}{!}{
    \begin{tabular}{lccccccccc}
    \toprule
    \multirow{2}{*}{Method} & \multicolumn{4}{c}{Overall Performance} & \multicolumn{5}{c}{LIVE-Qualcomm~\cite{ghadiyaram2018capture}} \\
    & SROCC$\uparrow$ & KROCC$\uparrow$ & PLCC$\uparrow$ & RMSE$\downarrow$ & SROCC$\uparrow$ & $p$-value (<0.05) & KROCC$\uparrow$ & PLCC$\uparrow$ & RMSE$\downarrow$ \\
    \midrule
    BRISQUE~\cite{mittal2012no} & 0.643 ($\pm$ 0.059) & 0.465 ($\pm$ 0.047) & 0.625 ($\pm$ 0.053) & 3.895 ($\pm$ 0.380) & 0.504 ($\pm$ 0.147) & 1.21E-04 & 0.365 ($\pm$ 0.111) & 0.516 ($\pm$ 0.127) & \underline{10.731} ($\pm$ 1.335) \\
    NIQE~\cite{mittal2013making} & 0.526 ($\pm$ 0.055) & 0.369 ($\pm$ 0.041) & 0.542 ($\pm$ 0.054) & 4.214 ($\pm$ 0.323) & 0.463 ($\pm$ 0.105) & 5.28E-07 & 0.328 ($\pm$ 0.088) & 0.464 ($\pm$ 0.136) & 10.858 ($\pm$ 1.013) \\
    CORNIA~\cite{ye2012unsupervised} &  0.591 ($\pm$ 0.052) & 0.423 ($\pm$ 0.043) & 0.595 ($\pm$ 0.051) & 4.139 ($\pm$ 0.300) & 0.460 ($\pm$ 0.130) & 4.98E-06 & 0.324 ($\pm$ 0.104) & 0.494 ($\pm$ 0.133) & 10.759 ($\pm$ 0.939) \\
    VIIDEO~\cite{mittal2016completely} & 0.237 ($\pm$ 0.073) & 0.164 ($\pm$ 0.050) & 0.218 ($\pm$ 0.070) & 5.115 ($\pm$ 0.285) & 0.127 ($\pm$ 0.137) & 9.77E-11 & 0.082 ($\pm$ 0.099) & -0.001 ($\pm$ 0.106) & 12.308 ($\pm$ 0.881) \\
    VBLIINDS~\cite{saad2014blind} & \underline{0.686} ($\pm$ 0.035) & \underline{0.503} ($\pm$ 0.032) & \underline{0.660} ($\pm$ 0.037) & \underline{3.753} ($\pm$ 0.365) & \underline{0.566} ($\pm$ 0.078) & 1.02E-05 & \underline{0.405} ($\pm$ 0.074) & \underline{0.568} ($\pm$ 0.089) & 10.760 ($\pm$ 1.231) \\
    \midrule
    \textbf{Ours} & \textbf{0.771} ($\pm$ 0.028) & \textbf{0.582} ($\pm$ 0.029) & \textbf{0.762} ($\pm$ 0.031) & \textbf{3.074} ($\pm$ 0.448) & \textbf{0.737} ($\pm$ 0.045) & - & \textbf{0.552} ($\pm$ 0.047) & \textbf{0.732} ($\pm$ 0.0360) & \textbf{8.863} ($\pm$ 1.042)
 \\
    \bottomrule
    \end{tabular}
    }

    \bigskip
    
    \resizebox{\textwidth}{!}{
    \begin{tabular}{lcccccccccc}
    \toprule
    \multirow{2}{*}{Method} & \multicolumn{5}{c}{KoNViD-1k~\cite{hosu2017konstanz}} & \multicolumn{5}{c}{CVD2014~\cite{nuutinen2016cvd2014}} \\
    & SROCC$\uparrow$ & $p$-value & KROCC$\uparrow$ & PLCC$\uparrow$ & RMSE$\downarrow$ & SROCC$\uparrow$ & $p$-value & KROCC$\uparrow$ & PLCC$\uparrow$ & RMSE$\downarrow$ \\
    \midrule
    BRISQUE~\cite{mittal2012no}  & 0.654 ($\pm$ 0.042) & 6.00E-06 & 0.473 ($\pm$ 0.034) & 0.626 ($\pm$ 0.041) & 0.507 ($\pm$ 0.031) & 0.709 ($\pm$ 0.067) & 7.03E-07 & 0.518 ($\pm$ 0.060) & 0.715 ($\pm$ 0.048) & 15.197 ($\pm$ 1.325) \\
    NIQE~\cite{mittal2013making} & 0.544 ($\pm$ 0.040) & 7.31E-11 & 0.379 ($\pm$ 0.029) & 0.546 ($\pm$ 0.038) & 0.536 ($\pm$ 0.010) & 0.489 ($\pm$ 0.091) & 1.73E-10 & 0.358 ($\pm$ 0.064) & 0.593 ($\pm$ 0.065) & 17.168 ($\pm$ 1.318) \\
    CORNIA~\cite{ye2012unsupervised} & 0.610 ($\pm$ 0.034) & 6.77E-09 & 0.436 ($\pm$ 0.029) & 0.608 ($\pm$ 0.032) & 0.509 ($\pm$ 0.014) & 0.614 ($\pm$ 0.075) & 5.69E-09 & 0.441 ($\pm$ 0.058) & 0.618 ($\pm$ 0.079) & 16.871 ($\pm$ 1.200) \\
    VIIDEO~\cite{mittal2016completely} & 0.298 ($\pm$ 0.052) & 4.22E-15 & 0.207 ($\pm$ 0.035) & 0.303 ($\pm$ 0.049) & 0.610 ($\pm$ 0.012) & 0.023 ($\pm$ 0.122) & 3.02E-14 & 0.021 ($\pm$ 0.081) & -0.025 ($\pm$ 0.144) & 21.822 ($\pm$ 1.152) \\
    VBLIINDS~\cite{saad2014blind} & \underline{0.695} ($\pm$ 0.024) & 6.75E-05 & \underline{0.509} ($\pm$ 0.020) & \underline{0.658} ($\pm$ 0.025) & \underline{0.483} ($\pm$ 0.011) & \underline{0.746} ($\pm$ 0.056) & 2.94E-06 & \underline{0.562} ($\pm$ 0.0570) & \underline{0.753} ($\pm$ 0.053) & \underline{14.292} ($\pm$ 1.413) \\
    \midrule
    \textbf{Ours} & \textbf{0.755} ($\pm$ 0.025) & - & \textbf{0.562} ($\pm$ 0.022) &	\textbf{0.744} ($\pm$ 0.029) & \textbf{0.469} ($\pm$ 0.054) & \textbf{0.880} ($\pm$ 0.030) & - & \textbf{0.705} ($\pm$ 0.044) & \textbf{0.885} ($\pm$ 0.031) & \textbf{11.287} ($\pm$ 1.943)\\
    \bottomrule
    \end{tabular}
    }
    \end{small}
\end{table*}

\subsection{Experimental Settings}
\indent\indent\textbf{Databases}. There are four databases constructed for our problem: LIVE Video Quality Challenge Database (LIVE-VQC)~\cite{sinno2018large}, Konstanz Natural Video Database (KoNViD-1k)~\cite{hosu2017konstanz}, LIVE-Qualcomm Mobile In-Capture Video Quality Database (LIVE-Qualcomm)~\cite{ghadiyaram2018capture}, and Camera Video Database (CVD2014)~\cite{nuutinen2016cvd2014}. The latter three are now publicly available, while the first one is not accessible now. So we conduct experiments on KoNViD-1k, LIVE-Qualcomm and CVD2014. Subjective quality scores are provided in the form of mean opinion score (MOS).

KoNViD-1k~\cite{hosu2017konstanz} aims at natural distortions. To guarantee the video content diversity, it comprises a total of 1,200 videos of resolution 960$\times$540 that are fairly sampled from a large public video dataset, YFCC100M. The videos are 8s with 24/25/30fps. The MOS ranges from 1.22 to 4.64.

LIVE-Qualcomm~\cite{ghadiyaram2018capture} aims at in-capture video distortions during video acquisition. It includes 208 videos of resolution 1920$\times$1080 captured by 8 different smart-phones and models 6 in-capture distortions (artifacts, color, exposure, focus, sharpness and stabilization). The videos are 15s with 30fps. The realignment MOS ranges from 16.5621 to 73.6428.

CVD2014~\cite{nuutinen2016cvd2014} also aims at complex distortions introduced during video acquisition. It contains 234 videos of resolution 640$\times$480 or 1280$\times$720 recorded by 78 different cameras. The videos are 10-25s with 11-31fps, which are a wide range of time span and fps. The realignment MOS ranges from -6.50 to 93.38.

\textbf{Compared methods}. Because only NR methods are applicable for quality assessment of in-the-wild videos, we choose five state-of-the-art NR methods (whose original codes are released by the authors) for comparison: VBLIINDS~\cite{saad2014blind}, VIIDEO~\cite{mittal2016completely}, BRISQUE~\cite{mittal2012no}\footnote{Video-level features of BRISQUE are the average pooling of its frame-level features.}, NIQE~\cite{mittal2013making}, and CORNIA~\cite{ye2012unsupervised}. Note that we cannot compare with the three recent deep learning-based general VQA methods, since \cite{zhang2018blind} needs scores of full-reference methods and \cite{kim2018deep,zhang2019objective} are full-reference methods, which are unfeasible for our problem.

\textbf{Basic evaluation criteria}. Spearman's rank-order correlation coefficient (SROCC), Kendall's rank-order correlation coefficient (KROCC), Pearson's linear correlation coefficient (PLCC) and root mean square error (RMSE) are the four performance criteria of VQA methods. SROCC and KROCC indicate the prediction monotonicity, while PLCC and RMSE measure the prediction accuracy. Better VQA methods should have larger SROCC/KROCC/PLCC and smaller RMSE. When the objective scores (\textit{i.e.},  the quality scores predicted by a VQA method) are not the same scale as the subjective scores, we refer to the suggestion of Video Quality Experts Group (VQEG)~\cite{vqeg2000fr} before calculating PLCC and RMSE values, and adopt a four-parameter logistic function for mapping the objective score $o$ to the subjective score $s$:
\begin{equation}\label{eq:nonlinear}
f(o)=\frac{\tau_1-\tau_2}{1+e^{-\frac{o-\tau_3}{\tau_4}}}+\tau_2,
\end{equation}
where $\tau_1$ to $\tau_4$ are fitting parameters initialized with $\tau_1=\max(s)$, $\tau_2=\min(s)$, $\tau_3=\mathrm{mean}(o)$, $\tau_4=\mathrm{std}(o)/4$.

\subsection{Performance Comparison}
\label{sec:performance}

For each database, 60\%, 20\%, and 20\% data are used for training, validation, and testing, respectively. There is no overlap among these three parts. This procedure is repeated 10 times and the mean and standard deviation of performance values are reported in Table~\ref{tab:performance}. For VBLIINDS, BRISQUE and our method, we choose the models with the highest SROCC values on the validation set during the training phase. NIQE, CORNIA, and VIIDEO are tested on the same 20\% testing data after the parameters in Eqn.~(\ref{eq:nonlinear}) are optimized with the training and validation data.

Table~\ref{tab:performance} summarizes the performance values on the three databases, and the overall performance values (indicated by the weighted performance values) as well. Our method achieves the best overall performance in terms of both the prediction monotonicity (SROCC, KROCC) and the prediction accuracy (PLCC, RMSE), and have a large gain over the second-best method VBLIINDS. VIIDEO fails because it is based only on temporal scene statistics and cannot model the complex distortions. 
For all individual databases, our method outperforms the other compared methods by a large margin. For example, compared to the second-best method VBLIINDS, in terms of SROCC, our method achieves 30.21\% improvements on LIVE-Qualcomm, 8.63\% improvements on KoNViD-1k and 17.96\% improvements on CVD2014.
Among the three databases, LIVE-Qualcomm is the most challenging one for the compared methods and our method---not only mean performance values are small but also standard deviation values for all methods are large. This verifies the statement in~\cite{ghadiyaram2018capture} that videos in LIVE-Qualcomm challenge both human viewers and objective VQA models.

\textbf{Statistical significance}. We further carry out the statistical significance test to see whether the results shown in~Table~\ref{tab:performance} are statistical significant or not. On each database, the paired t-test is conducted at 5\% significance level using the SROCC values (in 10 runs) of our method and of the compared one. The $p$-values are shown in Table~\ref{tab:performance}. All are smaller than 0.05 and prove our method is significantly better than all the other five state-of-the-art methods.

\subsection{Ablation Study}
\label{sec:ablation}

\begin{figure}[!htb]
\begin{center}
   \subfloat[KoNViD-1k]{\includegraphics[width=.96\linewidth]{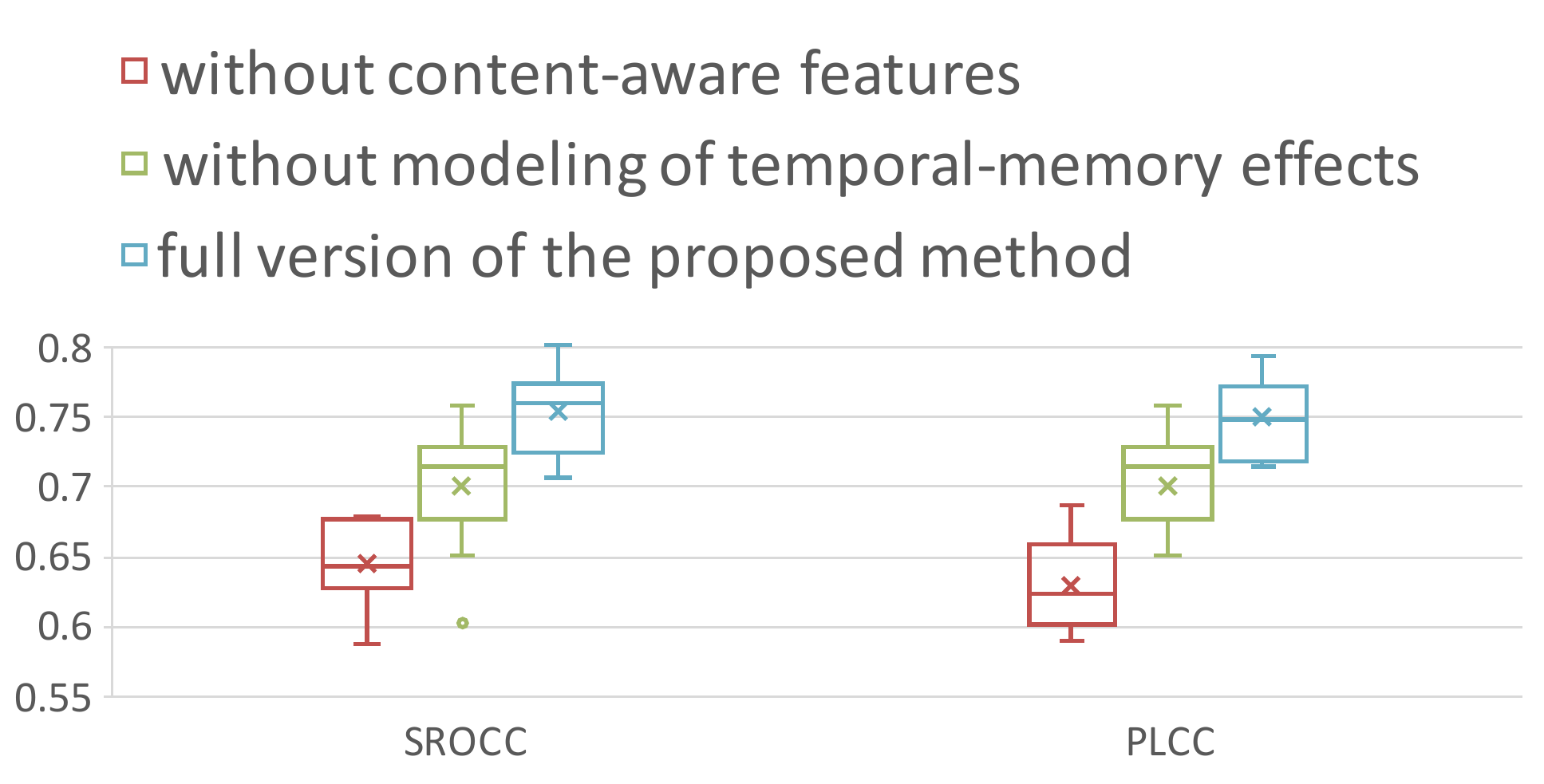}%
\label{fig:K}}
\hfil
   \subfloat[CVD2014]{\includegraphics[width=.96\linewidth]{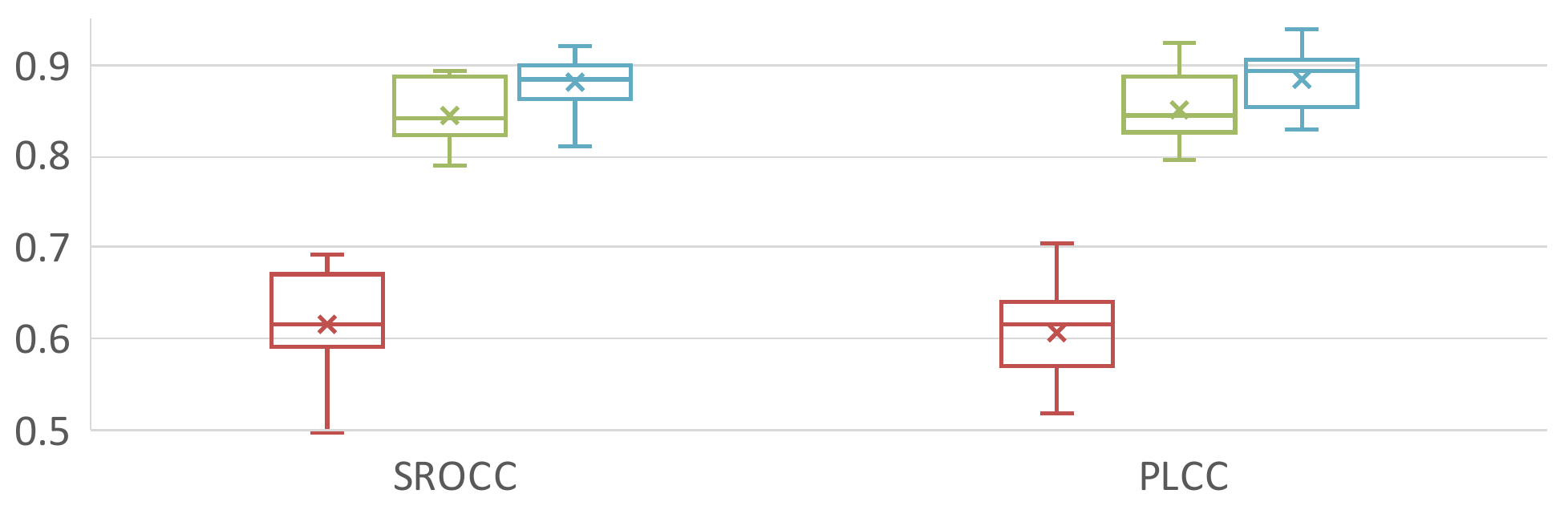}%
\label{fig:C}}
\hfil
   \subfloat[LIVE-Qualcomm]{\includegraphics[width=.96\linewidth]{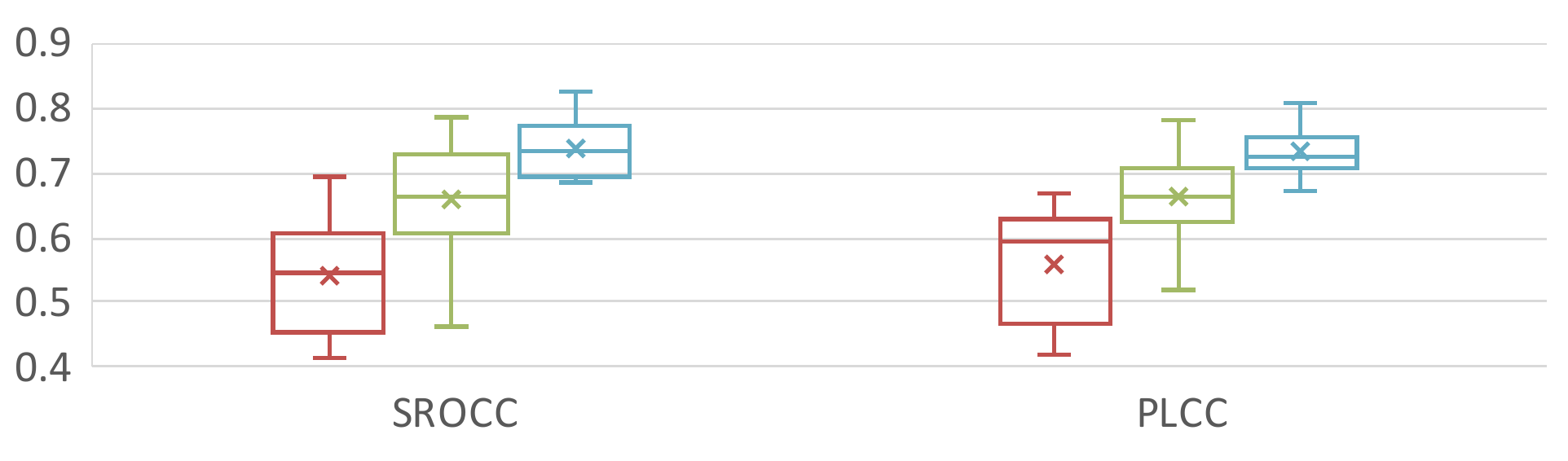}%
\label{fig:L}}
\end{center}
   \caption{Box plots of the ablation study.}
\label{fig:ablation}
\end{figure}

To demonstrate the importance of each module in our framework, we conduct an ablation study. The overall 10-run-results are shown in the form of box plots in Figure~\ref{fig:ablation}.

\textbf{Content-aware features}. We first show the performance drop due to the removal of the content-aware features. When we remove the content-aware features extracted from CNN, we use BRISQUE~\cite{mittal2012no} features instead ({\color[RGB]{179,87,81}red}). The removal of the content-aware features causes significant performance drop in all three databases. $p$-values are 1.10E-05, 1.76E-08, 2.47E-06, and 14.57\%, 30.00\%, 26.87\% decrease in terms of SROCC are found on KoNViD-1k, CVD2014 and LIVE-Qualcomm respectively. Content-aware perceptual features contribute most to our method, which verifies that content-aware perceptual features are crucial for assessing the perceived quality of in-the-wild videos.

\textbf{Modeling of temporal-memory effects}. To verify the effectiveness of modeling of temporal-memory effects, we compare the full version of our proposed method ({\color[RGB]{100,170,195}blue}) with the whole temporal modeling module removed ({\color[RGB]{162,185,102}green}). Temporal modeling provides 7.70\%, 4.14\%, 12.01\% SROCC gains on KoNViD-1k, CVD2014 and LIVE-Qualcomm respectively, where the $p$-values are 4.00E-04, 1.11E-04, and 8.49E-03. In view of PLCC, it leads to 5.98\%, 4.00\%, 10.41\% performance improvements on KoNViD-1k, CVD2014 and LIVE-Qualcomm respectively. We further do the ablation study on KoNViD-1k for the two individual temporal sub-modules separately. Removal of long-term dependencies modeling leads to 2.12\% decrease in terms of SROCC, while removal of subjectively-inspired temporal pooling leads to 2.68\% decrease in terms of SROCC. This indicates the two temporal sub-modules (one is global and the other is local) are complementary.

\subsection{Choice of Feature Extractor}
There are many choices for content-aware feature extraction. In the following, we mainly consider the pre-trained image classification models and the global standard deviation (std) pooling.

\textbf{Pre-trained image classification models}. In our implementation, we choose ResNet-50 as the content-aware feature extractor. It is interesting to explore other pre-trained image classification models for feature extraction. The results in Table~\ref{tab:cnn} show that VGG16 have similar performance with ResNet-50 (p-values of paired t-test using SROCC values are greater than 0.05, actually 0.1011). However, ResNet-50 has less parameters than AlexNet and VGG16.

\begin{table}[!hbt]
    \centering
    \caption{Performance of different pre-trained image classification models on KoNViD-1k.}
    \label{tab:cnn}
    \begin{small}
    \begin{tabular}{lccc}
    \toprule
    Pre-trained model & SROCC$\uparrow$ & KROCC$\uparrow$ & PLCC$\uparrow$ \\
    \midrule
    ResNet-50 & 0.755 ($\pm$0.025) & 0.562 ($\pm$0.022) & 0.744 ($\pm$0.029) \\
    AlexNet & 0.732 ($\pm$0.040) & 0.540 ($\pm$0.036) & 0.731 ($\pm$0.035) \\
    VGG16 & 0.745 ($\pm$0.024) & 0.554 ($\pm$0.023) & 0.747 ($\pm$0.022) \\
    \bottomrule
    \end{tabular}
    \end{small}
\end{table}

\begin{figure}[!htb]
\begin{center}
  \includegraphics[width=.9\linewidth]{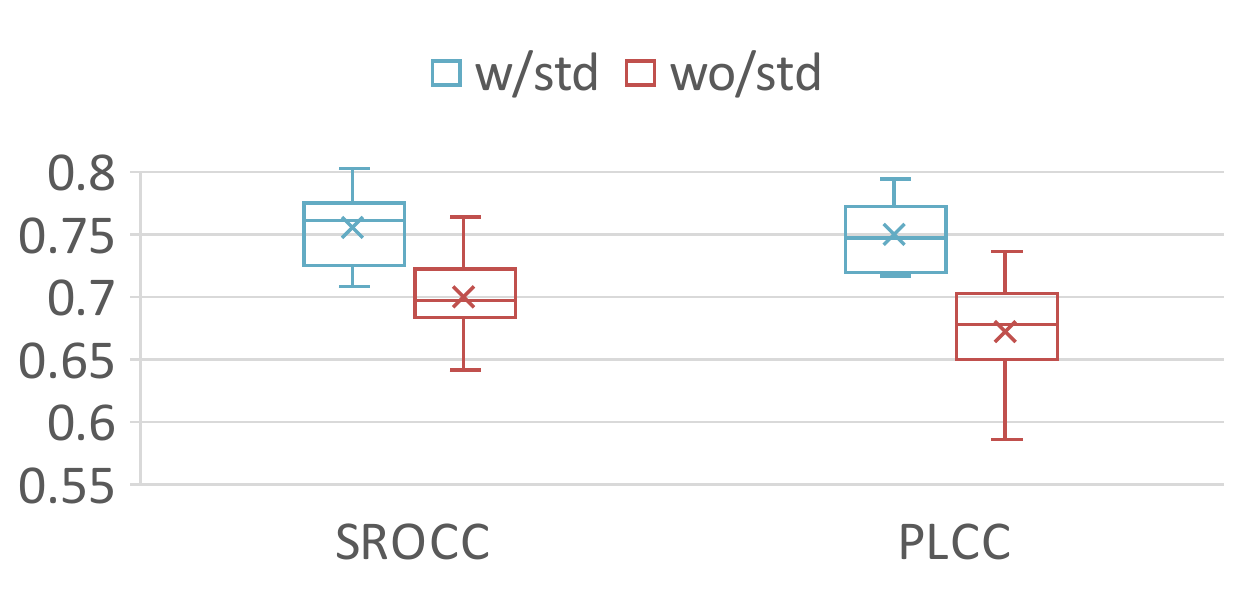}
\end{center}
   \caption{Effectiveness of global std pooling on KoNViD-1k.}
\label{fig:std}
\end{figure}

\textbf{Global std pooling}. When the global std pooling is removed, the performance on KoNViD-1k drops as shown in Figure~\ref{fig:std}. mean SROCC drops from 0.755 to 0.701, while mean PLCC drops significantly from 0.744 to 0.672. This verifies that global std pooling preserves more information and thus results in good performance.

\subsection{Choices of Temporal Pooling Strategy}
Here, we explore different choices of temporal pooling strategy.

\begin{figure}[!htb]
\begin{center}
  \includegraphics[width=\linewidth]{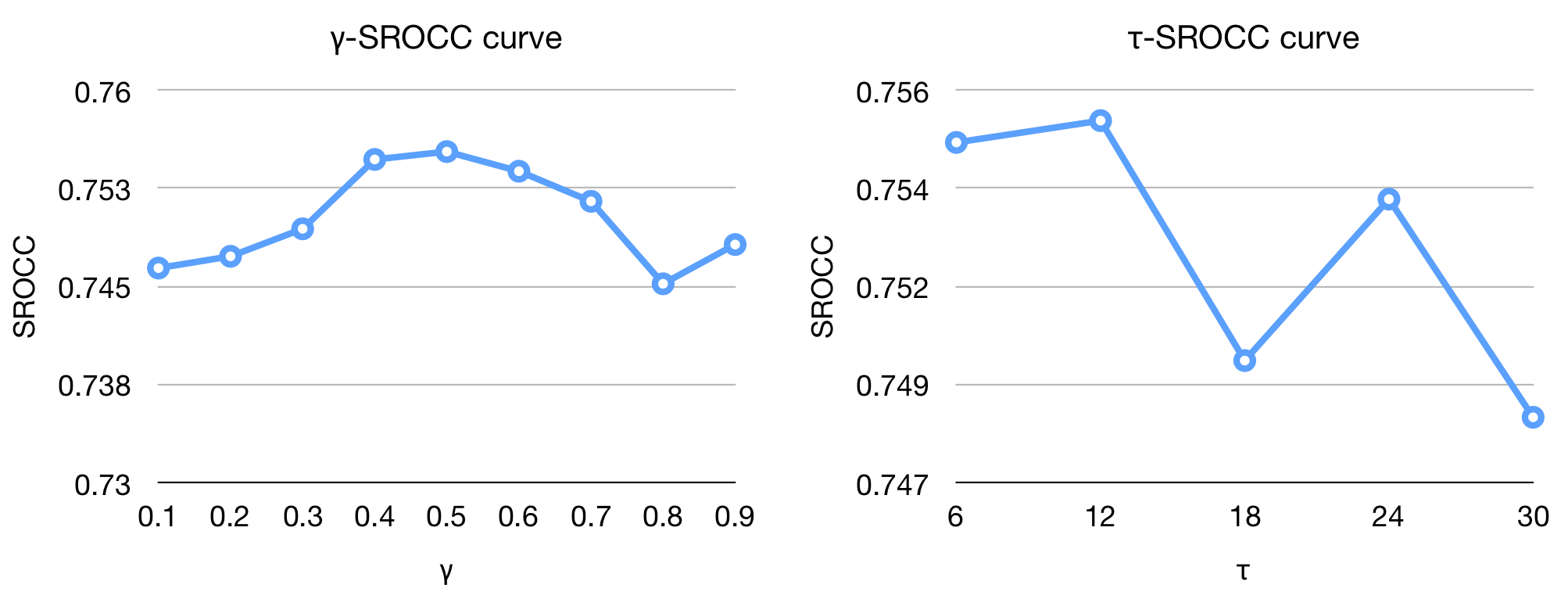}
\end{center}
   \caption{Performance on KoNViD-1k of different hyper-parameters in subjectively-inspired temporal pooling}
\label{fig:hyper-params}
\end{figure}

\textbf{Hyper-parameters in subjectively-inspired temporal pooling}. The subjectively-inspired temporal pooling contains two hyper-parameters, $\tau$ and $\gamma$. Figure~\ref{fig:hyper-params} shows results of different choices of the two parameters. In the left figure, $\tau$ is fixed to 12, and $\gamma$ varies from 0.1 to 0.9 with a step size 0.1. SROCC fluctuates up and down around 0.75, and achieves the best with $\gamma=0.5$. This is because smaller $\gamma$ overlooks the memory quality while larger $\gamma$ overlooks the current quality. In the right figure, $\gamma$ is fixed to 0.5, and $\tau$ varies from 6 to 30 with a step size 6. The highest SROCC value is obtained with $\tau=12$, which suggests temporal hysteresis effect may lasts about one second for videos with a frame rate of 25fps.

\textbf{Pooling in subjective-inspired temporal pooling}. To verify the effectiveness of min pooling, we compare it with average pooling. The results on KoNViD-1k are shown in Table~\ref{tab:min-pooling}. And we can see that average pooling is statistically worse than min pooling ($p$-value is 3.04E-04). This makes sense since min pooling accounts for ``humans are quick to criticize and slow to forgive''.

\begin{table}[!hbt]
    \centering
    \caption{Effectiveness of min pooling in subjective-inspired temporal pooling on KoNViD-1k.}
    \label{tab:min-pooling}
    \begin{small}
    \begin{tabular}{lcccc}
    \toprule
    pooling & SROCC$\uparrow$ & $p$-value & KROCC$\uparrow$ & PLCC$\uparrow$ \\
    \midrule
    min & \textbf{0.755} ($\pm$0.025) & - & \textbf{0.562} ($\pm$0.022) & \textbf{0.744} ($\pm$0.029) \\
    average & 0.736 ($\pm$0.031) & 3.04E-4 & 0.543 ($\pm$0.027) & 0.740 ($\pm$0.027)  \\
    \bottomrule
    \end{tabular}
    \end{small}
    
\end{table}

\textbf{Handcrafted weights vs. learned weights}.
Our subjectively-inspired temporal pooling can be regarded as a weighted average pooling strategy, where the weights are designed by hand (see Eqn.~(\ref{eq:memory score}), (\ref{eq:current score}) and (\ref{eq:approximate quality})) to mimic the temporal-memory effects. One interesting question is whether the performance can be further improved by making the weights learnable. One possible way is using a temporal CNN (TCNN) to learn the approximate scores $\mathbf{q}'$ from the frame quality scores $\mathbf{q}$, \textit{i.e.},
\begin{equation*}
    \mathbf{q}' = \mbox{TCNN}(\mathbf{q}, \mathbf{kernel\_size}=2\tau+1)=\mathbf{w}\otimes\mathbf{q},
\end{equation*}
where $\otimes$ means the convolutional operator, and $\mathbf{w}$ is the learnable weights of TCNN with length $2\tau+1$ (the same size as ours). 

Another way is by the convolutional neural aggregation network (CNAN) introduced in~\cite{kim2018deep}.  It is formulated as follow:
\begin{equation*}
    \bm{\omega} = \mathrm{softmax}(\mathbf{w}_m\otimes\mathbf{q}),\ Q =\bm{\omega}^T\mathbf{q},
\end{equation*}
where $\mathbf{w}_m$ is a memory kernel, $\bm{\omega}$ is the learned frame weights normalized by a softmax function and $Q$ is the overall video quality.

\begin{figure}[!htb]
\begin{center}
  \includegraphics[width=\linewidth]{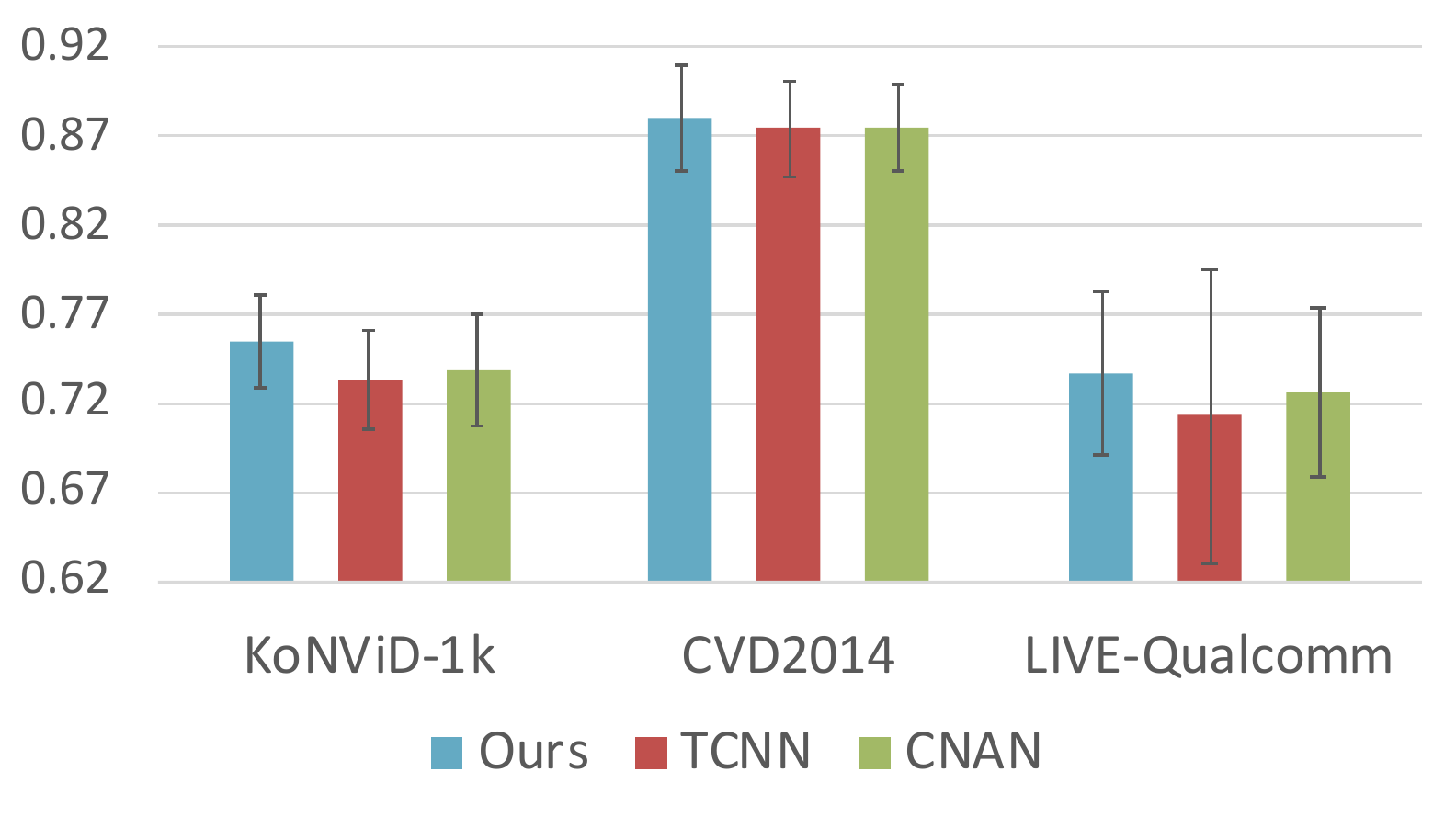}
\end{center}
   \caption{SROCC comparison between temporal pooling models with learned weights or handcrafted weights.}
\label{fig:weights}
\end{figure}

In Figure~\ref{fig:weights}, we report the mean and standard deviation of SROCC values among these three temporal pooling models (including ours) on the three databases. It can be seen that the two models with the learned weights (TCNN and CNAN) underperform the model with handcrafted weights (Ours). This may be explained by the fact that the handcrafted weights are manually designed to mimic the temporal hysteresis effects, while the learned weights do not capture the patterns well.

\subsection{Motion information}
\begin{figure}[!htb]
\begin{center}
   \includegraphics[width=.7\linewidth]{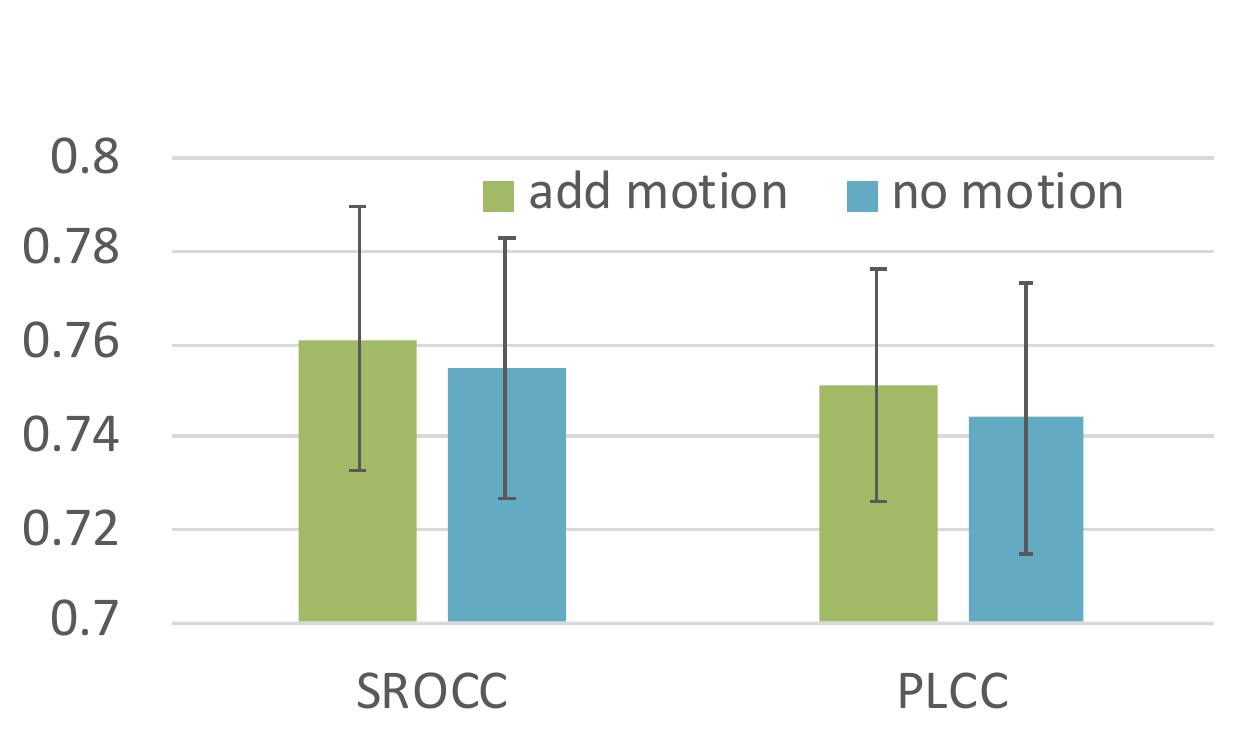}
\end{center}
   \caption{The performance comparison of our model with/without motion information on KoNViD-1k.}
\label{fig:motion}
\end{figure}

Motion information is important for video processing. In this subsection, we would like to see whether the performance can be further improved with the motion information added. We extract the optical flow using the initialized TVNet~\cite{fan2018end} without finetuning, and calculate the optical flow statistics as described in~\cite{manasa2016optical-NR}, then concatenate the statistics to the content-aware features. The performance comparison of our model with/without motion information on KoNViD-1k is shown in Figure~\ref{fig:motion}. Motion information can further improve the performance a little. However, we should note that optical flow computation is very expensive, which makes the small improvements seem unnecessary. It is desired to explore effective and efficient motion-aware features in the VQA task.

\subsection{Computational efficiency}
Besides the performance, computational efficiency is also crucial for NR-VQA methods. To provide a fair comparison for the computational efficiency of different methods, all tests are carried out on a desktop computer with Intel Core i7-6700K CPU@4.00 GHz, 12G NVIDIA TITAN Xp GPU and 64 GB RAM. The operating system is Ubuntu 14.04. The compared methods are implemented with MATLAB R2016b while our method is implemented with Python 3.6. The default settings of the original codes are used without any modification. From the three databases, we select four videos with different lengths and different resolutions for test. We repeat the tests ten times and the average computation time (seconds) for each method is shown in Table~\ref{tab:time}. Our method is faster than VBLIINDS---the method with the second-best performance. It is worth mentioning that our method can be accelerated to \textbf{30x faster or more} (The larger resolution is, the faster acceleration is.) by simply switching the CPU mode to the GPU mode.

\begin{table}[!hbt]
    \centering
    \caption{The average computation time (seconds) for four videos selected from the original databases. \{xxx\}frs@\{yyy\}p indicates the video frame length and the resolution.}
    \label{tab:time}
    \begin{small}
    \resizebox{\columnwidth}{!}{
    \begin{tabular}{lcccc}
    \toprule
    Method & 240frs@540p & 364frs@480p & 467frs@720p & 450frs@1080p \\
    \midrule
    BRISQUE~\cite{mittal2012no} & 12.6931 & 12.3405 & 41.2220 & 79.8119 \\
    NIQE~\cite{mittal2013making} & 45.6477 & 41.9705 & 155.9052 & 351.8327
 \\
    CORNIA~\cite{ye2012unsupervised} & 225.2185 & 325.5718 & 494.2449 & 616.4856 \\
    VIIDEO~\cite{mittal2016completely} & 137.0538 & 128.0868 & 465.2284 & 1024.5400 \\
    VBLIINDS~\cite{saad2014blind} & 382.0657 & 361.3868 & 1390.9999 & 3037.2960 \\
    Ours & 269.8371 & 249.2085 & 936.8452 & 2081.8400 \\
    \bottomrule
    \end{tabular}
    }
    \end{small}
\end{table}

\section{Conclusion and Future Work}
\label{sec:conclusion}
In this work, we propose a novel NR-VQA method for in-the-wild videos by incorporating two eminent effects of HVS, \textit{i.e.}, content-dependency and temporal-memory effects. 
Our proposed method is compared with five state-of-the-art methods on three publicly available in-the-wild VQA databases (KoNViD-1k, CVD2014, and LIVE-Qualcomm), and achieves 30.21\%, 8.63\%, and 17.96\% SROCC improvements on LIVE-Qualcomm, KoNViD-1k, and CVD2014, respectively.
Experiments also show that content-aware perceptual features and modeling of temporal-memory effects are of importance for in-the-wild video quality assessment.
However, the correlation values of the best method are still less than 0.76 on KoNViD-1k and LIVE-Qualcomm.
This indicates that there is ample room for developing an objective model which correlates well with human perception.
In the further study, we will consider embedding the spatio-temporal attention models into our framework since they could provide information about when and where the video is important for the VQA problem.

\section*{ACKNOWLEDGMENTS}
This work was partially supported by the National Basic Research Program of China (973 Program) under contract 2015CB351803, the Natural Science Foundation of China under contracts 61572042,  61520106004,  and 61527804. We acknowledge the High-Performance Computing Platform of Peking University for providing computational resources. 

\bibliographystyle{ACM-Reference-Format}
\bibliography{VSFA}


\begin{thebibliography}{57}


\ifx \showCODEN    \undefined \def \showCODEN     #1{\unskip}     \fi
\ifx \showDOI      \undefined \def \showDOI       #1{#1}\fi
\ifx \showISBNx    \undefined \def \showISBNx     #1{\unskip}     \fi
\ifx \showISBNxiii \undefined \def \showISBNxiii  #1{\unskip}     \fi
\ifx \showISSN     \undefined \def \showISSN      #1{\unskip}     \fi
\ifx \showLCCN     \undefined \def \showLCCN      #1{\unskip}     \fi
\ifx \shownote     \undefined \def \shownote      #1{#1}          \fi
\ifx \showarticletitle \undefined \def \showarticletitle #1{#1}   \fi
\ifx \showURL      \undefined \def \showURL       {\relax}        \fi
\providecommand\bibfield[2]{#2}
\providecommand\bibinfo[2]{#2}
\providecommand\natexlab[1]{#1}
\providecommand\showeprint[2][]{arXiv:#2}

\bibitem[\protect\citeauthoryear{Bampis, Li, Moorthy, Katsavounidis, Aaron, and
  Bovik}{Bampis et~al\mbox{.}}{2017}]%
        {bampis2017study}
\bibfield{author}{\bibinfo{person}{Christos~George Bampis},
  \bibinfo{person}{Zhi Li}, \bibinfo{person}{Anush~Krishna Moorthy},
  \bibinfo{person}{Ioannis Katsavounidis}, \bibinfo{person}{Anne Aaron}, {and}
  \bibinfo{person}{Alan~Conrad Bovik}.} \bibinfo{year}{2017}\natexlab{}.
\newblock \showarticletitle{Study of temporal effects on subjective video
  quality of experience}.
\newblock \bibinfo{journal}{\emph{TIP}} \bibinfo{volume}{26},
  \bibinfo{number}{11} (\bibinfo{date}{Nov.} \bibinfo{year}{2017}),
  \bibinfo{pages}{5217--5231}.
\newblock


\bibitem[\protect\citeauthoryear{Cho, Van~Merri{\"e}nboer, Gulcehre, Bahdanau,
  Bougares, Schwenk, and Bengio}{Cho et~al\mbox{.}}{2014}]%
        {cho2014learning}
\bibfield{author}{\bibinfo{person}{Kyunghyun Cho}, \bibinfo{person}{Bart
  Van~Merri{\"e}nboer}, \bibinfo{person}{Caglar Gulcehre},
  \bibinfo{person}{Dzmitry Bahdanau}, \bibinfo{person}{Fethi Bougares},
  \bibinfo{person}{Holger Schwenk}, {and} \bibinfo{person}{Yoshua Bengio}.}
  \bibinfo{year}{2014}\natexlab{}.
\newblock \showarticletitle{Learning phrase representations using {RNN}
  encoder-decoder for statistical machine translation}.
\newblock \bibinfo{journal}{\emph{arXiv preprint arXiv:1406.1078}}
  (\bibinfo{year}{2014}).
\newblock


\bibitem[\protect\citeauthoryear{Choi and Bovik}{Choi and Bovik}{2018}]%
        {choi2018video}
\bibfield{author}{\bibinfo{person}{Lark~Kwon Choi} {and}
  \bibinfo{person}{Alan~Conrad Bovik}.} \bibinfo{year}{2018}\natexlab{}.
\newblock \showarticletitle{Video quality assessment accounting for temporal
  visual masking of local flicker}.
\newblock \bibinfo{journal}{\emph{SPIC}}  \bibinfo{volume}{67}
  (\bibinfo{year}{2018}), \bibinfo{pages}{182--198}.
\newblock


\bibitem[\protect\citeauthoryear{Deng, Dong, Socher, Li, Li, and Fei-Fei}{Deng
  et~al\mbox{.}}{2009}]%
        {deng2009imagenet}
\bibfield{author}{\bibinfo{person}{Jia Deng}, \bibinfo{person}{Wei Dong},
  \bibinfo{person}{Richard Socher}, \bibinfo{person}{Li-Jia Li},
  \bibinfo{person}{Kai Li}, {and} \bibinfo{person}{Li Fei-Fei}.}
  \bibinfo{year}{2009}\natexlab{}.
\newblock \showarticletitle{{ImageNet}: A large-scale hierarchical image
  database}. In \bibinfo{booktitle}{\emph{CVPR}}. IEEE,
  \bibinfo{pages}{248--255}.
\newblock


\bibitem[\protect\citeauthoryear{Dodge and Karam}{Dodge and Karam}{2016}]%
        {dodge2016understanding}
\bibfield{author}{\bibinfo{person}{Samuel Dodge} {and} \bibinfo{person}{Lina
  Karam}.} \bibinfo{year}{2016}\natexlab{}.
\newblock \showarticletitle{Understanding how image quality affects deep neural
  networks}. In \bibinfo{booktitle}{\emph{QoMEX}}. IEEE.
\newblock


\bibitem[\protect\citeauthoryear{Duanmu, Ma, and Wang}{Duanmu
  et~al\mbox{.}}{2017}]%
        {duanmu2017quality}
\bibfield{author}{\bibinfo{person}{Zhengfang Duanmu}, \bibinfo{person}{Kede
  Ma}, {and} \bibinfo{person}{Zhou Wang}.} \bibinfo{year}{2017}\natexlab{}.
\newblock \showarticletitle{Quality-of-experience of adaptive video streaming:
  Exploring the space of adaptations}. In \bibinfo{booktitle}{\emph{ACM MM}}.
  ACM, \bibinfo{pages}{1752--1760}.
\newblock


\bibitem[\protect\citeauthoryear{Fan, Huang, Chuang~Gan, Gong, and Huang}{Fan
  et~al\mbox{.}}{2018}]%
        {fan2018end}
\bibfield{author}{\bibinfo{person}{Lijie Fan}, \bibinfo{person}{Wenbing Huang},
  \bibinfo{person}{Stefano~Ermon Chuang~Gan}, \bibinfo{person}{Boqing Gong},
  {and} \bibinfo{person}{Junzhou Huang}.} \bibinfo{year}{2018}\natexlab{}.
\newblock \showarticletitle{End-to-End Learning of Motion Representation for
  Video Understanding}. In \bibinfo{booktitle}{\emph{CVPR}}.
  \bibinfo{pages}{6016--6025}.
\newblock


\bibitem[\protect\citeauthoryear{Freitas, Akamine, and Farias}{Freitas
  et~al\mbox{.}}{2018}]%
        {freitas2018using}
\bibfield{author}{\bibinfo{person}{Pedro~Garcia Freitas},
  \bibinfo{person}{Welington~YL Akamine}, {and} \bibinfo{person}{Myl{\`e}ne~CQ
  Farias}.} \bibinfo{year}{2018}\natexlab{}.
\newblock \showarticletitle{Using multiple spatio-temporal features to estimate
  video quality}.
\newblock \bibinfo{journal}{\emph{SPIC}}  \bibinfo{volume}{64}
  (\bibinfo{year}{2018}), \bibinfo{pages}{1--10}.
\newblock


\bibitem[\protect\citeauthoryear{Ghadiyaram, Chen, Inguva, and
  Kokaram}{Ghadiyaram et~al\mbox{.}}{2017}]%
        {ghadiyaram2017no}
\bibfield{author}{\bibinfo{person}{Deepti Ghadiyaram}, \bibinfo{person}{Chao
  Chen}, \bibinfo{person}{Sasi Inguva}, {and} \bibinfo{person}{Anil Kokaram}.}
  \bibinfo{year}{2017}\natexlab{}.
\newblock \showarticletitle{A no-reference video quality predictor for
  compression and scaling artifacts}. In \bibinfo{booktitle}{\emph{ICIP}}.
  IEEE, \bibinfo{pages}{3445--3449}.
\newblock


\bibitem[\protect\citeauthoryear{Ghadiyaram, Pan, Bovik, Moorthy, Panda, and
  Yang}{Ghadiyaram et~al\mbox{.}}{2018}]%
        {ghadiyaram2018capture}
\bibfield{author}{\bibinfo{person}{Deepti Ghadiyaram}, \bibinfo{person}{Janice
  Pan}, \bibinfo{person}{Alan~C Bovik}, \bibinfo{person}{Anush~K Moorthy},
  \bibinfo{person}{Prasanjit Panda}, {and} \bibinfo{person}{Kai-Chieh Yang}.}
  \bibinfo{year}{2018}\natexlab{}.
\newblock \showarticletitle{In-Capture mobile video distortions: A study of
  subjective behavior and objective algorithms}.
\newblock \bibinfo{journal}{\emph{IEEE TCSVT}} \bibinfo{volume}{28},
  \bibinfo{number}{9} (\bibinfo{date}{Sept.} \bibinfo{year}{2018}),
  \bibinfo{pages}{2061--2077}.
\newblock


\bibitem[\protect\citeauthoryear{He, Zhang, Ren, and Sun}{He
  et~al\mbox{.}}{2016}]%
        {he2016deep}
\bibfield{author}{\bibinfo{person}{Kaiming He}, \bibinfo{person}{Xiangyu
  Zhang}, \bibinfo{person}{Shaoqing Ren}, {and} \bibinfo{person}{Jian Sun}.}
  \bibinfo{year}{2016}\natexlab{}.
\newblock \showarticletitle{Deep residual learning for image recognition}. In
  \bibinfo{booktitle}{\emph{CVPR}}. \bibinfo{pages}{770--778}.
\newblock


\bibitem[\protect\citeauthoryear{Hosu, Hahn, Jenadeleh, Lin, Men, Szir{\'a}nyi,
  Li, and Saupe}{Hosu et~al\mbox{.}}{2017}]%
        {hosu2017konstanz}
\bibfield{author}{\bibinfo{person}{Vlad Hosu}, \bibinfo{person}{Franz Hahn},
  \bibinfo{person}{Mohsen Jenadeleh}, \bibinfo{person}{Hanhe Lin},
  \bibinfo{person}{Hui Men}, \bibinfo{person}{Tam{\'a}s Szir{\'a}nyi},
  \bibinfo{person}{Shujun Li}, {and} \bibinfo{person}{Dietmar Saupe}.}
  \bibinfo{year}{2017}\natexlab{}.
\newblock \showarticletitle{{The Konstanz natural video database (KoNViD-1k)}}.
  In \bibinfo{booktitle}{\emph{QoMEX}}. IEEE.
\newblock


\bibitem[\protect\citeauthoryear{Jaramillo, Ni{\~n}o-Casta{\~n}eda,
  Plati{\v{s}}a, and Philips}{Jaramillo et~al\mbox{.}}{2016}]%
        {jaramillo2016content}
\bibfield{author}{\bibinfo{person}{Benhur~Ortiz Jaramillo},
  \bibinfo{person}{Jorge~Oswaldo Ni{\~n}o-Casta{\~n}eda},
  \bibinfo{person}{Ljiljana Plati{\v{s}}a}, {and} \bibinfo{person}{Wilfried
  Philips}.} \bibinfo{year}{2016}\natexlab{}.
\newblock \showarticletitle{Content-aware objective video quality assessment}.
\newblock \bibinfo{journal}{\emph{JEI}} \bibinfo{volume}{25},
  \bibinfo{number}{1} (\bibinfo{year}{2016}), \bibinfo{pages}{013011}.
\newblock


\bibitem[\protect\citeauthoryear{Juluri, Tamarapalli, and Medhi}{Juluri
  et~al\mbox{.}}{2015}]%
        {juluri2015measurement}
\bibfield{author}{\bibinfo{person}{Parikshit Juluri},
  \bibinfo{person}{Venkatesh Tamarapalli}, {and} \bibinfo{person}{Deep Medhi}.}
  \bibinfo{year}{2015}\natexlab{}.
\newblock \showarticletitle{Measurement of quality of experience of
  video-on-demand services: A survey}.
\newblock \bibinfo{journal}{\emph{IEEE Commun. Surv. Tutor.}}
  \bibinfo{volume}{18}, \bibinfo{number}{1} (\bibinfo{year}{2015}),
  \bibinfo{pages}{401--418}.
\newblock


\bibitem[\protect\citeauthoryear{Kim, Kim, Ahn, Kim, and Lee}{Kim
  et~al\mbox{.}}{2018}]%
        {kim2018deep}
\bibfield{author}{\bibinfo{person}{Woojae Kim}, \bibinfo{person}{Jongyoo Kim},
  \bibinfo{person}{Sewoong Ahn}, \bibinfo{person}{Jinwoo Kim}, {and}
  \bibinfo{person}{Sanghoon Lee}.} \bibinfo{year}{2018}\natexlab{}.
\newblock \showarticletitle{Deep Video Quality Assessor: From Spatio-temporal
  Visual Sensitivity to A Convolutional Neural Aggregation Network}. In
  \bibinfo{booktitle}{\emph{ECCV}}. \bibinfo{pages}{219--234}.
\newblock


\bibitem[\protect\citeauthoryear{Kingma and Ba}{Kingma and Ba}{2014}]%
        {kingma2014adam}
\bibfield{author}{\bibinfo{person}{Diederik~P Kingma} {and}
  \bibinfo{person}{Jimmy Ba}.} \bibinfo{year}{2014}\natexlab{}.
\newblock \showarticletitle{{Adam}: A method for stochastic optimization}.
\newblock \bibinfo{journal}{\emph{arXiv preprint arXiv:1412.6980}}
  (\bibinfo{year}{2014}).
\newblock


\bibitem[\protect\citeauthoryear{Li, Jiang, Lin, and Jiang}{Li
  et~al\mbox{.}}{2019}]%
        {li2019has}
\bibfield{author}{\bibinfo{person}{Dingquan Li}, \bibinfo{person}{Tingting
  Jiang}, \bibinfo{person}{Weisi Lin}, {and} \bibinfo{person}{Ming Jiang}.}
  \bibinfo{year}{2019}\natexlab{}.
\newblock \showarticletitle{Which Has Better Visual Quality: The Clear Blue Sky
  or a Blurry Animal?}
\newblock \bibinfo{journal}{\emph{IEEE TMM}} \bibinfo{volume}{21},
  \bibinfo{number}{5} (\bibinfo{date}{May} \bibinfo{year}{2019}),
  \bibinfo{pages}{1221--1234}.
\newblock


\bibitem[\protect\citeauthoryear{Li, Guo, and Lu}{Li et~al\mbox{.}}{2016a}]%
        {li2016spatiotemporal}
\bibfield{author}{\bibinfo{person}{Xuelong Li}, \bibinfo{person}{Qun Guo},
  {and} \bibinfo{person}{Xiaoqiang Lu}.} \bibinfo{year}{2016}\natexlab{a}.
\newblock \showarticletitle{Spatiotemporal statistics for video quality
  assessment}.
\newblock \bibinfo{journal}{\emph{IEEE TIP}} \bibinfo{volume}{25},
  \bibinfo{number}{7} (\bibinfo{date}{July} \bibinfo{year}{2016}),
  \bibinfo{pages}{3329--3342}.
\newblock


\bibitem[\protect\citeauthoryear{Li, Po, Cheung, Xu, Feng, Yuan, and Cheung}{Li
  et~al\mbox{.}}{2016b}]%
        {li2016no}
\bibfield{author}{\bibinfo{person}{Yuming Li}, \bibinfo{person}{Lai-Man Po},
  \bibinfo{person}{Chun-Ho Cheung}, \bibinfo{person}{Xuyuan Xu},
  \bibinfo{person}{Litong Feng}, \bibinfo{person}{Fang Yuan}, {and}
  \bibinfo{person}{Kwok-Wai Cheung}.} \bibinfo{year}{2016}\natexlab{b}.
\newblock \showarticletitle{No-reference video quality assessment with {3D}
  shearlet transform and convolutional neural networks}.
\newblock \bibinfo{journal}{\emph{IEEE TCSVT}} \bibinfo{volume}{26},
  \bibinfo{number}{6} (\bibinfo{date}{June} \bibinfo{year}{2016}),
  \bibinfo{pages}{1044--1057}.
\newblock


\bibitem[\protect\citeauthoryear{Liu, Duanmu, and Wang}{Liu
  et~al\mbox{.}}{2018}]%
        {liu2018end}
\bibfield{author}{\bibinfo{person}{Wentao Liu}, \bibinfo{person}{Zhengfang
  Duanmu}, {and} \bibinfo{person}{Zhou Wang}.} \bibinfo{year}{2018}\natexlab{}.
\newblock \showarticletitle{End-to-End Blind Quality Assessment of Compressed
  Videos Using Deep Neural Networks}. In \bibinfo{booktitle}{\emph{ACM MM}}.
  ACM, \bibinfo{pages}{546--554}.
\newblock


\bibitem[\protect\citeauthoryear{Lu, He, Yang, Jia, and Gao}{Lu
  et~al\mbox{.}}{2019}]%
        {lu2019spatiotemporal}
\bibfield{author}{\bibinfo{person}{Wen Lu}, \bibinfo{person}{Ran He},
  \bibinfo{person}{Jiachen Yang}, \bibinfo{person}{Changcheng Jia}, {and}
  \bibinfo{person}{Xinbo Gao}.} \bibinfo{year}{2019}\natexlab{}.
\newblock \showarticletitle{A spatiotemporal model of video quality assessment
  via {3D} gradient differencing}.
\newblock \bibinfo{journal}{\emph{Information Sciences}}  \bibinfo{volume}{478}
  (\bibinfo{year}{2019}), \bibinfo{pages}{141--151}.
\newblock


\bibitem[\protect\citeauthoryear{Manasa and Channappayya}{Manasa and
  Channappayya}{2016}]%
        {manasa2016optical-NR}
\bibfield{author}{\bibinfo{person}{K Manasa} {and} \bibinfo{person}{Sumohana~S
  Channappayya}.} \bibinfo{year}{2016}\natexlab{}.
\newblock \showarticletitle{An optical flow-based no-reference video quality
  assessment algorithm}. In \bibinfo{booktitle}{\emph{ICIP}}. IEEE,
  \bibinfo{pages}{2400--2404}.
\newblock


\bibitem[\protect\citeauthoryear{Men, Lin, and Saupe}{Men
  et~al\mbox{.}}{2017}]%
        {men2017empirical}
\bibfield{author}{\bibinfo{person}{Hui Men}, \bibinfo{person}{Hanhe Lin}, {and}
  \bibinfo{person}{Dietmar Saupe}.} \bibinfo{year}{2017}\natexlab{}.
\newblock \showarticletitle{Empirical evaluation of no-reference {VQA} methods
  on a natural video quality database}. In \bibinfo{booktitle}{\emph{QoMEX}}.
  IEEE.
\newblock


\bibitem[\protect\citeauthoryear{Men, Lin, and Saupe}{Men
  et~al\mbox{.}}{2018}]%
        {men2018spatiotemporal}
\bibfield{author}{\bibinfo{person}{Hui Men}, \bibinfo{person}{Hanhe Lin}, {and}
  \bibinfo{person}{Dietmar Saupe}.} \bibinfo{year}{2018}\natexlab{}.
\newblock \showarticletitle{Spatiotemporal Feature Combination Model for
  No-Reference Video Quality Assessment}. In \bibinfo{booktitle}{\emph{QoMEX}}.
  IEEE.
\newblock


\bibitem[\protect\citeauthoryear{Mikhailiuk, P{\'e}rez-Ortiz, and
  Mantiuk}{Mikhailiuk et~al\mbox{.}}{2018}]%
        {mikhailiuk2018psychometric}
\bibfield{author}{\bibinfo{person}{Aliaksei Mikhailiuk},
  \bibinfo{person}{Mar{\'\i}a P{\'e}rez-Ortiz}, {and} \bibinfo{person}{Rafal
  Mantiuk}.} \bibinfo{year}{2018}\natexlab{}.
\newblock \showarticletitle{Psychometric scaling of {TID2013} dataset}. In
  \bibinfo{booktitle}{\emph{QoMEX}}.
\newblock


\bibitem[\protect\citeauthoryear{Mirkovic, Vrgovic, Culibrk, Stefanovic, and
  Anderla}{Mirkovic et~al\mbox{.}}{2014}]%
        {mirkovic2014evaluating}
\bibfield{author}{\bibinfo{person}{Milan Mirkovic}, \bibinfo{person}{Petar
  Vrgovic}, \bibinfo{person}{Dubravko Culibrk}, \bibinfo{person}{Darko
  Stefanovic}, {and} \bibinfo{person}{Andras Anderla}.}
  \bibinfo{year}{2014}\natexlab{}.
\newblock \showarticletitle{Evaluating the role of content in subjective video
  quality assessment}.
\newblock \bibinfo{journal}{\emph{The Scientific World Journal}}
  \bibinfo{volume}{2014} (\bibinfo{year}{2014}).
\newblock


\bibitem[\protect\citeauthoryear{Mittal, Moorthy, and Bovik}{Mittal
  et~al\mbox{.}}{2012}]%
        {mittal2012no}
\bibfield{author}{\bibinfo{person}{Anish Mittal},
  \bibinfo{person}{Anush~Krishna Moorthy}, {and} \bibinfo{person}{Alan~Conrad
  Bovik}.} \bibinfo{year}{2012}\natexlab{}.
\newblock \showarticletitle{No-reference image quality assessment in the
  spatial domain}.
\newblock \bibinfo{journal}{\emph{IEEE TIP}} \bibinfo{volume}{21},
  \bibinfo{number}{12} (\bibinfo{date}{Dec.} \bibinfo{year}{2012}),
  \bibinfo{pages}{4695--4708}.
\newblock


\bibitem[\protect\citeauthoryear{Mittal, Saad, and Bovik}{Mittal
  et~al\mbox{.}}{2016}]%
        {mittal2016completely}
\bibfield{author}{\bibinfo{person}{Anish Mittal}, \bibinfo{person}{Michele~A
  Saad}, {and} \bibinfo{person}{Alan~C Bovik}.}
  \bibinfo{year}{2016}\natexlab{}.
\newblock \showarticletitle{A completely blind video integrity oracle}.
\newblock \bibinfo{journal}{\emph{IEEE TIP}} \bibinfo{volume}{25},
  \bibinfo{number}{1} (\bibinfo{date}{Jan.} \bibinfo{year}{2016}),
  \bibinfo{pages}{289--300}.
\newblock


\bibitem[\protect\citeauthoryear{Mittal, Soundararajan, and Bovik}{Mittal
  et~al\mbox{.}}{2013}]%
        {mittal2013making}
\bibfield{author}{\bibinfo{person}{Anish Mittal}, \bibinfo{person}{Rajiv
  Soundararajan}, {and} \bibinfo{person}{Alan~C Bovik}.}
  \bibinfo{year}{2013}\natexlab{}.
\newblock \showarticletitle{Making a ``completely blind" image quality
  analyzer}.
\newblock \bibinfo{journal}{\emph{IEEE SPL}} \bibinfo{volume}{20},
  \bibinfo{number}{3} (\bibinfo{date}{Mar.} \bibinfo{year}{2013}),
  \bibinfo{pages}{209--212}.
\newblock


\bibitem[\protect\citeauthoryear{Moorthy, Choi, Bovik, and De~Veciana}{Moorthy
  et~al\mbox{.}}{2012}]%
        {moorthy2012video}
\bibfield{author}{\bibinfo{person}{Anush~Krishna Moorthy},
  \bibinfo{person}{Lark~Kwon Choi}, \bibinfo{person}{Alan~Conrad Bovik}, {and}
  \bibinfo{person}{Gustavo De~Veciana}.} \bibinfo{year}{2012}\natexlab{}.
\newblock \showarticletitle{Video quality assessment on mobile devices:
  Subjective, behavioral and objective studies}.
\newblock \bibinfo{journal}{\emph{IEEE JSTSP}} \bibinfo{volume}{6},
  \bibinfo{number}{6} (\bibinfo{date}{Oct.} \bibinfo{year}{2012}),
  \bibinfo{pages}{652--671}.
\newblock


\bibitem[\protect\citeauthoryear{Nuutinen, Virtanen, Vaahteranoksa, Vuori,
  Oittinen, and H{\"a}kkinen}{Nuutinen et~al\mbox{.}}{2016}]%
        {nuutinen2016cvd2014}
\bibfield{author}{\bibinfo{person}{Mikko Nuutinen}, \bibinfo{person}{Toni
  Virtanen}, \bibinfo{person}{Mikko Vaahteranoksa}, \bibinfo{person}{Tero
  Vuori}, \bibinfo{person}{Pirkko Oittinen}, {and} \bibinfo{person}{Jukka
  H{\"a}kkinen}.} \bibinfo{year}{2016}\natexlab{}.
\newblock \showarticletitle{{CVD2014}---A database for evaluating no-reference
  video quality assessment algorithms}.
\newblock \bibinfo{journal}{\emph{IEEE TIP}} \bibinfo{volume}{25},
  \bibinfo{number}{7} (\bibinfo{date}{July} \bibinfo{year}{2016}),
  \bibinfo{pages}{3073--3086}.
\newblock


\bibitem[\protect\citeauthoryear{Park, Seshadrinathan, Lee, and Bovik}{Park
  et~al\mbox{.}}{2013}]%
        {park2013video}
\bibfield{author}{\bibinfo{person}{Jincheol Park}, \bibinfo{person}{Kalpana
  Seshadrinathan}, \bibinfo{person}{Sanghoon Lee}, {and}
  \bibinfo{person}{Alan~Conrad Bovik}.} \bibinfo{year}{2013}\natexlab{}.
\newblock \showarticletitle{Video quality pooling adaptive to perceptual
  distortion severity}.
\newblock \bibinfo{journal}{\emph{IEEE TIP}} \bibinfo{volume}{22},
  \bibinfo{number}{2} (\bibinfo{date}{Feb.} \bibinfo{year}{2013}),
  \bibinfo{pages}{610--620}.
\newblock


\bibitem[\protect\citeauthoryear{Paszke, Gross, Chintala, Chanan, Yang, DeVito,
  Lin, Desmaison, Antiga, and Lerer}{Paszke et~al\mbox{.}}{2017}]%
        {paszke2017automatic}
\bibfield{author}{\bibinfo{person}{Adam Paszke}, \bibinfo{person}{Sam Gross},
  \bibinfo{person}{Soumith Chintala}, \bibinfo{person}{Gregory Chanan},
  \bibinfo{person}{Edward Yang}, \bibinfo{person}{Zachary DeVito},
  \bibinfo{person}{Zeming Lin}, \bibinfo{person}{Alban Desmaison},
  \bibinfo{person}{Luca Antiga}, {and} \bibinfo{person}{Adam Lerer}.}
  \bibinfo{year}{2017}\natexlab{}.
\newblock \showarticletitle{Automatic differentiation in {PyTorch}}. In
  \bibinfo{booktitle}{\emph{NIPS-W}}.
\newblock


\bibitem[\protect\citeauthoryear{Rimac-Drlje, Vranjes, and Zagar}{Rimac-Drlje
  et~al\mbox{.}}{2009}]%
        {rimac2009influence}
\bibfield{author}{\bibinfo{person}{Snjezana Rimac-Drlje},
  \bibinfo{person}{Mario Vranjes}, {and} \bibinfo{person}{Drago Zagar}.}
  \bibinfo{year}{2009}\natexlab{}.
\newblock \showarticletitle{Influence of temporal pooling method on the
  objective video quality evaluation}. In \bibinfo{booktitle}{\emph{BMSB}}.
  IEEE, \bibinfo{pages}{1--5}.
\newblock


\bibitem[\protect\citeauthoryear{Saad, Bovik, and Charrier}{Saad
  et~al\mbox{.}}{2014}]%
        {saad2014blind}
\bibfield{author}{\bibinfo{person}{Michele~A Saad}, \bibinfo{person}{Alan~C
  Bovik}, {and} \bibinfo{person}{Christophe Charrier}.}
  \bibinfo{year}{2014}\natexlab{}.
\newblock \showarticletitle{Blind prediction of natural video quality}.
\newblock \bibinfo{journal}{\emph{IEEE TIP}} \bibinfo{volume}{23},
  \bibinfo{number}{3} (\bibinfo{date}{Mar.} \bibinfo{year}{2014}),
  \bibinfo{pages}{1352--1365}.
\newblock


\bibitem[\protect\citeauthoryear{Seshadrinathan and Bovik}{Seshadrinathan and
  Bovik}{2010}]%
        {seshadrinathan2010motion}
\bibfield{author}{\bibinfo{person}{Kalpana Seshadrinathan} {and}
  \bibinfo{person}{Alan~Conrad Bovik}.} \bibinfo{year}{2010}\natexlab{}.
\newblock \showarticletitle{Motion tuned spatio-temporal quality assessment of
  natural videos}.
\newblock \bibinfo{journal}{\emph{IEEE TIP}} \bibinfo{volume}{19},
  \bibinfo{number}{2} (\bibinfo{date}{Feb.} \bibinfo{year}{2010}),
  \bibinfo{pages}{335--350}.
\newblock


\bibitem[\protect\citeauthoryear{Seshadrinathan and Bovik}{Seshadrinathan and
  Bovik}{2011}]%
        {seshadrinathan2011temporal}
\bibfield{author}{\bibinfo{person}{Kalpana Seshadrinathan} {and}
  \bibinfo{person}{Alan~C Bovik}.} \bibinfo{year}{2011}\natexlab{}.
\newblock \showarticletitle{Temporal hysteresis model of time varying
  subjective video quality}. In \bibinfo{booktitle}{\emph{ICASSP}}. IEEE,
  \bibinfo{pages}{1153--1156}.
\newblock


\bibitem[\protect\citeauthoryear{Seshadrinathan, Soundararajan, Bovik, and
  Cormack}{Seshadrinathan et~al\mbox{.}}{2010}]%
        {seshadrinathan2010study}
\bibfield{author}{\bibinfo{person}{Kalpana Seshadrinathan},
  \bibinfo{person}{Rajiv Soundararajan}, \bibinfo{person}{Alan~Conrad Bovik},
  {and} \bibinfo{person}{Lawrence~K Cormack}.} \bibinfo{year}{2010}\natexlab{}.
\newblock \showarticletitle{Study of subjective and objective quality
  assessment of video}.
\newblock \bibinfo{journal}{\emph{IEEE TIP}} \bibinfo{volume}{19},
  \bibinfo{number}{6} (\bibinfo{date}{June} \bibinfo{year}{2010}),
  \bibinfo{pages}{1427--1441}.
\newblock


\bibitem[\protect\citeauthoryear{Seufert, Egger, Slanina, Zinner, Ho{\ss}feld,
  and Tran-Gia}{Seufert et~al\mbox{.}}{2014}]%
        {seufert2014survey}
\bibfield{author}{\bibinfo{person}{Michael Seufert}, \bibinfo{person}{Sebastian
  Egger}, \bibinfo{person}{Martin Slanina}, \bibinfo{person}{Thomas Zinner},
  \bibinfo{person}{Tobias Ho{\ss}feld}, {and} \bibinfo{person}{Phuoc
  Tran-Gia}.} \bibinfo{year}{2014}\natexlab{}.
\newblock \showarticletitle{A survey on quality of experience of {HTTP}
  adaptive streaming}.
\newblock \bibinfo{journal}{\emph{IEEE Commun. Surv. Tutor.}}
  \bibinfo{volume}{17}, \bibinfo{number}{1} (\bibinfo{year}{2014}),
  \bibinfo{pages}{469--492}.
\newblock


\bibitem[\protect\citeauthoryear{Seufert, Slanina, Egger, and Kottkamp}{Seufert
  et~al\mbox{.}}{2013}]%
        {seufert2013pool}
\bibfield{author}{\bibinfo{person}{Michael Seufert}, \bibinfo{person}{Martin
  Slanina}, \bibinfo{person}{Sebastian Egger}, {and} \bibinfo{person}{Meik
  Kottkamp}.} \bibinfo{year}{2013}\natexlab{}.
\newblock \showarticletitle{``To pool or not to pool": A comparison of temporal
  pooling methods for HTTP adaptive video streaming}. In
  \bibinfo{booktitle}{\emph{QoMEX}}. IEEE, \bibinfo{pages}{52--57}.
\newblock


\bibitem[\protect\citeauthoryear{Siahaan, Hanjalic, and Redi}{Siahaan
  et~al\mbox{.}}{2018}]%
        {siahaan2018semantic}
\bibfield{author}{\bibinfo{person}{Ernestasia Siahaan}, \bibinfo{person}{Alan
  Hanjalic}, {and} \bibinfo{person}{Judith~A Redi}.}
  \bibinfo{year}{2018}\natexlab{}.
\newblock \showarticletitle{Semantic-aware blind image quality assessment}.
\newblock \bibinfo{journal}{\emph{SPIC}}  \bibinfo{volume}{60}
  (\bibinfo{year}{2018}), \bibinfo{pages}{237--252}.
\newblock


\bibitem[\protect\citeauthoryear{Sinno and Bovik}{Sinno and Bovik}{2019}]%
        {sinno2018large}
\bibfield{author}{\bibinfo{person}{Zeina Sinno} {and} \bibinfo{person}{Alan~C
  Bovik}.} \bibinfo{year}{2019}\natexlab{}.
\newblock \showarticletitle{Large scale study of perceptual video quality}.
\newblock \bibinfo{journal}{\emph{IEEE TIP}} \bibinfo{volume}{28},
  \bibinfo{number}{2} (\bibinfo{date}{Feb.} \bibinfo{year}{2019}),
  \bibinfo{pages}{612--627}.
\newblock


\bibitem[\protect\citeauthoryear{Triantaphillidou, Allen, and
  Jacobson}{Triantaphillidou et~al\mbox{.}}{2007}]%
        {triantaphillidou2007image}
\bibfield{author}{\bibinfo{person}{Sophie Triantaphillidou},
  \bibinfo{person}{Elizabeth Allen}, {and} \bibinfo{person}{R Jacobson}.}
  \bibinfo{year}{2007}\natexlab{}.
\newblock \showarticletitle{{Image quality comparison between JPEG and
  JPEG2000. II. Scene dependency, scene analysis, and classification}}.
\newblock \bibinfo{journal}{\emph{JIST}} \bibinfo{volume}{51},
  \bibinfo{number}{3} (\bibinfo{year}{2007}), \bibinfo{pages}{259--270}.
\newblock


\bibitem[\protect\citeauthoryear{VQEG}{VQEG}{2000}]%
        {vqeg2000fr}
\bibfield{author}{\bibinfo{person}{VQEG}.} \bibinfo{year}{2000}\natexlab{}.
\newblock \bibinfo{title}{Final report from the {Video Quality Experts Group}
  on the validation of objective models of video quality assessment}.
\newblock
\newblock


\bibitem[\protect\citeauthoryear{Vu and Chandler}{Vu and Chandler}{2014}]%
        {vu2014vis3}
\bibfield{author}{\bibinfo{person}{Phong~V Vu} {and} \bibinfo{person}{Damon~M
  Chandler}.} \bibinfo{year}{2014}\natexlab{}.
\newblock \showarticletitle{{ViS$_3$}: An algorithm for video quality
  assessment via analysis of spatial and spatiotemporal slices}.
\newblock \bibinfo{journal}{\emph{JEI}} \bibinfo{volume}{23},
  \bibinfo{number}{1} (\bibinfo{year}{2014}), \bibinfo{pages}{013016}.
\newblock


\bibitem[\protect\citeauthoryear{Wang, Katsavounidis, Zhou, Park, Lei, Zhou,
  Pun, Jin, Wang, Wang, Zhang, Huang, Kwong, and Kuo}{Wang
  et~al\mbox{.}}{2017}]%
        {wang2017videoset}
\bibfield{author}{\bibinfo{person}{Haiqiang Wang}, \bibinfo{person}{Ioannis
  Katsavounidis}, \bibinfo{person}{Jiantong Zhou}, \bibinfo{person}{Jeonghoon
  Park}, \bibinfo{person}{Shawmin Lei}, \bibinfo{person}{Xin Zhou},
  \bibinfo{person}{Man-On Pun}, \bibinfo{person}{Xin Jin},
  \bibinfo{person}{Ronggang Wang}, \bibinfo{person}{Xu Wang},
  \bibinfo{person}{Yun Zhang}, \bibinfo{person}{Jiwu Huang},
  \bibinfo{person}{Sam Kwong}, {and} \bibinfo{person}{C.-C.~Jay Kuo}.}
  \bibinfo{year}{2017}\natexlab{}.
\newblock \showarticletitle{{VideoSet}: A large-scale compressed video quality
  dataset based on {JND} measurement}.
\newblock \bibinfo{journal}{\emph{JVCIR}}  \bibinfo{volume}{46}
  (\bibinfo{year}{2017}), \bibinfo{pages}{292--302}.
\newblock


\bibitem[\protect\citeauthoryear{Wang, Jiang, Ma, and Gao}{Wang
  et~al\mbox{.}}{2012}]%
        {wang2012novel}
\bibfield{author}{\bibinfo{person}{Yue Wang}, \bibinfo{person}{Tingting Jiang},
  \bibinfo{person}{Siwei Ma}, {and} \bibinfo{person}{Wen Gao}.}
  \bibinfo{year}{2012}\natexlab{}.
\newblock \showarticletitle{Novel spatio-temporal structural information based
  video quality metric}.
\newblock \bibinfo{journal}{\emph{IEEE TCSVT}} \bibinfo{volume}{22},
  \bibinfo{number}{7} (\bibinfo{date}{July} \bibinfo{year}{2012}),
  \bibinfo{pages}{989--998}.
\newblock


\bibitem[\protect\citeauthoryear{Wang, Lu, and Bovik}{Wang
  et~al\mbox{.}}{2004}]%
        {wang2004video}
\bibfield{author}{\bibinfo{person}{Zhou Wang}, \bibinfo{person}{Ligang Lu},
  {and} \bibinfo{person}{Alan~C Bovik}.} \bibinfo{year}{2004}\natexlab{}.
\newblock \showarticletitle{Video quality assessment based on structural
  distortion measurement}.
\newblock \bibinfo{journal}{\emph{SPIC}} \bibinfo{volume}{19},
  \bibinfo{number}{2} (\bibinfo{year}{2004}), \bibinfo{pages}{121--132}.
\newblock


\bibitem[\protect\citeauthoryear{Wu, Zeng, Dong, Shi, and Lin}{Wu
  et~al\mbox{.}}{2019}]%
        {wu2019blind}
\bibfield{author}{\bibinfo{person}{Jinjian Wu}, \bibinfo{person}{Jichen Zeng},
  \bibinfo{person}{Weisheng Dong}, \bibinfo{person}{Guangming Shi}, {and}
  \bibinfo{person}{Weisi Lin}.} \bibinfo{year}{2019}\natexlab{}.
\newblock \showarticletitle{Blind image quality assessment with hierarchy:
  Degradation from local structure to deep semantics}.
\newblock \bibinfo{journal}{\emph{JVCIR}}  \bibinfo{volume}{58}
  (\bibinfo{year}{2019}), \bibinfo{pages}{353--362}.
\newblock


\bibitem[\protect\citeauthoryear{Xu, Ye, Liu, and Doermann}{Xu
  et~al\mbox{.}}{2014}]%
        {xu2014no}
\bibfield{author}{\bibinfo{person}{Jingtao Xu}, \bibinfo{person}{Peng Ye},
  \bibinfo{person}{Yong Liu}, {and} \bibinfo{person}{David Doermann}.}
  \bibinfo{year}{2014}\natexlab{}.
\newblock \showarticletitle{No-reference video quality assessment via feature
  learning}. In \bibinfo{booktitle}{\emph{ICIP}}. IEEE,
  \bibinfo{pages}{491--495}.
\newblock


\bibitem[\protect\citeauthoryear{Ye, Kumar, Kang, and Doermann}{Ye
  et~al\mbox{.}}{2012}]%
        {ye2012unsupervised}
\bibfield{author}{\bibinfo{person}{Peng Ye}, \bibinfo{person}{Jayant Kumar},
  \bibinfo{person}{Le Kang}, {and} \bibinfo{person}{David Doermann}.}
  \bibinfo{year}{2012}\natexlab{}.
\newblock \showarticletitle{Unsupervised feature learning framework for
  no-reference image quality assessment}. In \bibinfo{booktitle}{\emph{CVPR}}.
  IEEE, \bibinfo{pages}{1098--1105}.
\newblock


\bibitem[\protect\citeauthoryear{You, Ebrahimi, and Perkis}{You
  et~al\mbox{.}}{2014}]%
        {you2014attention}
\bibfield{author}{\bibinfo{person}{Junyong You}, \bibinfo{person}{Touradj
  Ebrahimi}, {and} \bibinfo{person}{Andrew Perkis}.}
  \bibinfo{year}{2014}\natexlab{}.
\newblock \showarticletitle{Attention driven foveated video quality
  assessment}.
\newblock \bibinfo{journal}{\emph{IEEE TIP}} \bibinfo{volume}{23},
  \bibinfo{number}{1} (\bibinfo{year}{2014}), \bibinfo{pages}{200--213}.
\newblock


\bibitem[\protect\citeauthoryear{Zhang, Isola, Efros, Shechtman, and
  Wang}{Zhang et~al\mbox{.}}{2018b}]%
        {zhang2018unreasonable}
\bibfield{author}{\bibinfo{person}{Richard Zhang}, \bibinfo{person}{Phillip
  Isola}, \bibinfo{person}{Alexei~A Efros}, \bibinfo{person}{Eli Shechtman},
  {and} \bibinfo{person}{Oliver Wang}.} \bibinfo{year}{2018}\natexlab{b}.
\newblock \showarticletitle{The unreasonable effectiveness of deep features as
  a perceptual metric}. In \bibinfo{booktitle}{\emph{CVPR}}.
  \bibinfo{pages}{586--595}.
\newblock


\bibitem[\protect\citeauthoryear{Zhang and Liu}{Zhang and Liu}{2017}]%
        {zhang2017study}
\bibfield{author}{\bibinfo{person}{Wei Zhang} {and} \bibinfo{person}{Hantao
  Liu}.} \bibinfo{year}{2017}\natexlab{}.
\newblock \showarticletitle{Study of saliency in objective video quality
  assessment}.
\newblock \bibinfo{journal}{\emph{IEEE TIP}} \bibinfo{volume}{26},
  \bibinfo{number}{3} (\bibinfo{date}{Mar.} \bibinfo{year}{2017}),
  \bibinfo{pages}{1275--1288}.
\newblock


\bibitem[\protect\citeauthoryear{Zhang, Gao, He, Lu, and He}{Zhang
  et~al\mbox{.}}{2018a}]%
        {zhang2018blind}
\bibfield{author}{\bibinfo{person}{Yu Zhang}, \bibinfo{person}{Xinbo Gao},
  \bibinfo{person}{Lihuo He}, \bibinfo{person}{Wen Lu}, {and}
  \bibinfo{person}{Ran He}.} \bibinfo{year}{2018}\natexlab{a}.
\newblock \showarticletitle{Blind Video Quality Assessment with Weakly
  Supervised Learning and Resampling Strategy}.
\newblock \bibinfo{journal}{\emph{TCSVT}} (\bibinfo{year}{2018}).
\newblock


\bibitem[\protect\citeauthoryear{Zhang, Gao, He, Lu, and He}{Zhang
  et~al\mbox{.}}{2019}]%
        {zhang2019objective}
\bibfield{author}{\bibinfo{person}{Yu Zhang}, \bibinfo{person}{Xinbo Gao},
  \bibinfo{person}{Lihuo He}, \bibinfo{person}{Wen Lu}, {and}
  \bibinfo{person}{Ran He}.} \bibinfo{year}{2019}\natexlab{}.
\newblock \showarticletitle{Objective Video Quality Assessment Combining
  Transfer Learning With {CNN}}.
\newblock \bibinfo{journal}{\emph{TNNLS}} (\bibinfo{year}{2019}).
\newblock


\bibitem[\protect\citeauthoryear{Zhu, Wang, and Shuai}{Zhu
  et~al\mbox{.}}{2017}]%
        {zhu2017blind}
\bibfield{author}{\bibinfo{person}{Yun Zhu}, \bibinfo{person}{Yongfang Wang},
  {and} \bibinfo{person}{Yuan Shuai}.} \bibinfo{year}{2017}\natexlab{}.
\newblock \showarticletitle{Blind video quality assessment based on
  spatio-temporal internal generative mechanism}. In
  \bibinfo{booktitle}{\emph{ICIP}}. IEEE, \bibinfo{pages}{305--309}.
\newblock


\end{thebibliography}

\end{document}